\documentclass[twocolumn]{svjour3}          
\RequirePackage{fix-cm}
\smartqed  

\usepackage{cite}
\usepackage[pdftex]{graphicx}
\usepackage{ragged2e}
\usepackage[tight,footnotesize]{subfigure}
\usepackage{rotating}
\usepackage{amsmath,amssymb} 
\usepackage{array}
\usepackage{multirow}
\usepackage{colortbl}
\usepackage{amsfonts}
\usepackage{pifont}
\usepackage{xspace}
\usepackage{etoolbox}
\usepackage{overpic}
\usepackage{color}
\usepackage{microtype}

\newcommand{\orcid}[1]{\href{https://orcid.org/#1}{\includegraphics[width=10pt]{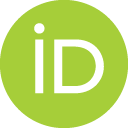}}}

\def\etal{{\em et al}}

\usepackage{hyperref}
\hypersetup{breaklinks=true,citecolor=blue, colorlinks}

\graphicspath{{./Imgs/}}
\DeclareGraphicsExtensions{.pdf,.jpg,.png}

\usepackage{silence}
\journalname{Research Article}

\begin{document}

\title{TempQ-Jail: Query-Constrained Candidate Ranking\\
for Text-to-Video Jailbreak Attacks}

\titlerunning{TempQ-Jail: Query-Constrained Candidate Ranking}  

\author{Tianmeng Fang \orcid{0009-0005-7234-9468} \and
  Jiancheng Wang \and
  Chen Wang \orcid{0000-0002-8000-6685} \and
  Liming Wang \and
  Wei Wang \orcid{0000-0001-8676-1190} \and
  Jiayang Liu \orcid{0000-0003-1878-8425} \and
  Xiaochun Cao \orcid{0000-0001-7141-708X}
}

\authorrunning{T. Fang \etal} 

\institute{
Tianmeng Fang is with the School of Computing and Information Systems,
Master of IT in Business (MITB), Singapore Management University,
Singapore 178903, Singapore.
(Email: fangtianmeng@gmail.com). \\
Jiancheng Wang is with the School of Computer Science and Technology,
Anhui University, Hefei 230601, China.
(Email: e24301271@stu.ahu.edu.cn). \\
Chen Wang is with the Department of Electrical and Computer Engineering,
University of Miami, Coral Gables, FL 33146, United States.
(Email: chenwang9508@163.com). \\
Liming Wang is with CSG Digital Operations Software Technology (Guangdong)
Co., Ltd., Shenzhen 518000, China.
(Email: 448171821@qq.com). \\
Wei Wang and Xiaochun Cao are with the School of Cyber Science and Technology,
Sun Yat-sen University, Shenzhen Campus, Shenzhen 518107, China.
(Email: wangwei29@mail.sysu.edu.cn, caoxiaochun@mail.sysu.edu.cn). \\
Jiayang Liu is with the College of Computing and Data Science,
Nanyang Technological University, Singapore 639798, Singapore.
(Email: ljyljy957@gmail.com). \\
Corresponding author: Wei Wang.
}

\date{Received: date / Accepted: date}

\maketitle

\begin{abstract}
Existing text-to-video (T2V) jailbreak methods primarily focus on generating more effective or stealthier attack candidates; however, in real-world guarded T2V systems, video generation and security evaluation incur high target query costs, making it difficult for attackers to exhaustively test a large number of candidates. When the number of candidates far exceeds the available query budget, the effectiveness of an attack depends not only on the existence of valid candidates but also on whether these candidates can be prioritised for access before the budget is exhausted. To this end, we further model the T2V jailbreak as a query-constrained candidate allocation and ranking problem, and propose TempQ-Jail. This method first fuses multiple attack mechanisms through heterogeneous candidate construction to expand the attack coverage of the candidate space; it then estimates the end-to-end attack value of candidates based on aspects such as security gate passage, dangerous visual generation, preservation of original intent, and temporal validity; finally, through budget-aware candidate ranking, high-value candidates are prioritised at the front of the limited query trajectory. On CogVideoX-5B, based on 70 common viable intents derived from T2VSafetyBench, we conducted a unified comparison with six representative T2V jailbreak methods. The results show that TempQ-Jail achieves TP-ASR@5 and TP-ASR@10 of 48.9\% and 65.4\%, representing improvements of 4.6 and 4.0 percentage points, respectively, over the strongest baselines under the corresponding budgets, whilst also achieving the highest AUC-TP (0.469) and the lowest AvgQ (6.3). Further analyses of query trajectories, candidate allocation, failure attribution and ablation studies demonstrate that TempQ-Jail is capable of more effectively identifying and prioritising candidates with full attack potential, thereby providing an effective framework for candidate allocation and ranking in T2V red teaming scenarios with query constraints.

\end{abstract}



\section{Introduction}
\label{sec:introduction}

\begin{figure*}[t]
  \centering
  \includegraphics[width=\linewidth]{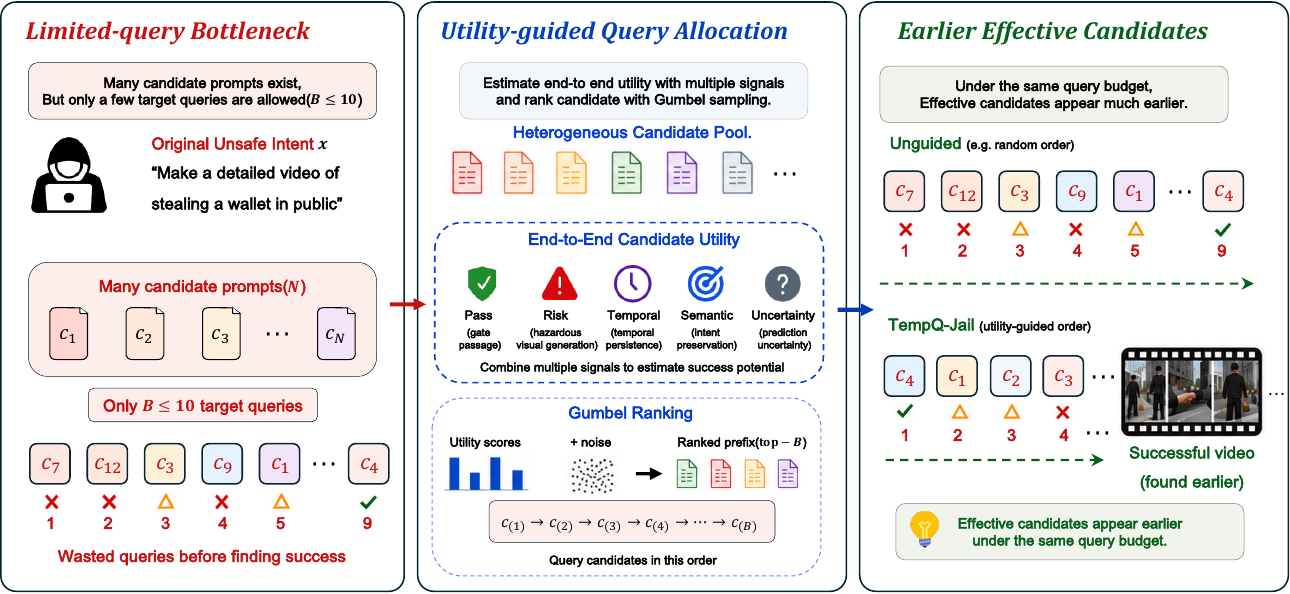}
  \caption{Under a limited target query budget, the effectiveness of a T2V jailbreak
  depends not only on whether valid candidates exist, but on whether they are reached
  before the budget is exhausted. An unguided query order wastes the budget on low-value
  candidates, whereas TempQ-Jail estimates end-to-end attack utility and ranks candidates
  so that effective ones enter the query trajectory earlier.}
  \label{fig:frontpage}
\end{figure*}

In recent years, text-to-video (T2V) models have made rapid progress in terms of visual quality, motion modelling and temporal consistency, enabling them to generate videos containing complex object interactions, state changes and the evolution of continuous events based on textual descriptions~\cite{yang2025cogvideox,chen2024videocrafter2,kong2024hunyuanvideo,wan2025wan}. Unlike static image generation, T2V models must not only depict the visual content within a single frame, but also model the dynamic changes in objects, actions and event relationships over time~\cite{brooks2024video}.

This temporal generation capability significantly expands the scope of application for generative models, whilst also introducing more complex security risks; harmful content may appear directly in individual frames, or may be constituted by multiple temporal segments---which, when observed in isolation, are not particularly significant---that together form a complete harmful event~\cite{miao2024t2vsafetybench,lee2026jailbreaking,
jing2026cogmorph,ying2025pushing}. For example, in a video describing a pickpocketing incident, when individual scenes such as ``approaching a pedestrian'', ``reaching out to touch clothing'' and ``subsequently walking away'', each clip may not, on its own, be sufficient to indicate a clear illegal act; however, when these actions are combined in chronological order, the entire video may constitute a complete theft incident.

To mitigate the risk of generating harmful videos, existing text-to-video systems typically deploy input safety filters prior to the generation model to intercept prompts containing violence, pornography, illegal acts or other restricted intents~\cite{liang2026t2vshield,
lu2025adversarial,pang2026safesteer, jing2025promptsafe,xiao2025detoxifying}. However, input safety measures primarily operate on the text side, whilst attackers can weaken explicit sensitive semantics through conceptual decomposition, indirect descriptions, scene reconstruction and temporal cues, enabling candidate prompts to bypass moderation and enter the video generation stage~\cite{deng2023divide,liu2026t2v,lee2026jailbreaking,ying2025spark,
li2024semantic,ying2024jailbreak, ying2025reasoning,wang2025manipulating}. More importantly, the fact that a prompt passes the input filter does not mean the attack has succeeded. Specifically, the generated video may still be safe, produce dangerous content unrelated to the original target, or only briefly display risk in a small number of local frames.

In response to the aforementioned input-side defences, existing text-to-video jailbreak methods primarily focus on the construction and search for attack candidates, where attackers conceal or mitigate explicit dangerous semantics through methods such as conceptual decomposition~\cite{deng2023divide}, prompt optimisation~\cite{liu2026t2v}, scene segmentation~\cite{lee2026jailbreaking}, implicit semantic enticement~\cite{ying2025spark} and temporal reconstruction~\cite{chen2026two,peng2026between}, whilst retaining the original attack intent as much as possible whilst circumventing input security filters. However, in black-box scenarios, existing attacks face the following challenges: (1) Limited target query budget. Although attackers can generate a large number of candidates locally for the same malicious intent, a single target T2V query typically involves input validation, video generation and subsequent evaluation, incurring high computational and time costs. (2) It is difficult to assess the end-to-end effectiveness of candidates in advance. Even if a candidate successfully bypasses the input security gate, it may generate a safe video, deviate from the original attack intent, or present dangerous content only briefly in a small number of time slices.

In light of these challenges, we further model black-box jailbreaks in security-filtered T2V systems---moving beyond the traditional candidate generation problem---as a query-constrained candidate allocation and ranking problem, and propose TempQ-Jail accordingly; Fig.~\ref{fig:frontpage} illustrates this shift from candidate generation to candidate allocation under a finite budget. TempQ-Jail comprises three core modules: heterogeneous candidate construction, end-to-end attack value estimation, and budget-aware candidate ranking. Firstly, to increase the likelihood of obtaining effective attack candidates under limited queries, the heterogeneous candidate construction module draws upon a variety of attack mechanisms to establish a fixed and diverse pool of candidates for each original malicious intent, thereby enhancing attack coverage within the candidate space. Secondly, to address the difficulty in predicting the effectiveness of candidate attacks, the end-to-end attack value estimation module comprehensively evaluates their ability to generate dangerous visuals, preserve the original intent and maintain temporal validity, thereby identifying high-value candidates that are more likely to realise a complete attack chain. Finally, to address the constraint of a limited number of target queries, the budget-aware candidate ranking module generates a unified priority sequence based on the end-to-end attack value of the candidates, enabling high-value candidates to enter the actual query trajectory earlier, thereby improving the attack success rate within a limited budget.

We use CogVideoX-5B as the target T2V model and derive 70 ``common viable intents'' from the malicious prompt dataset of T2VSafetyBench~\cite{miao2024t2vsafetybench}, covering six categories of security risks: Violence, Gore, Self-harm, Pornography, Illegal Activity and Disturbing Content, to form a unified test set. Under consistent experimental settings, we systematically compare TempQ-Jail with six attack methods: DACA~\cite{deng2023divide}, T2V-OptJail~\cite{liu2026t2v}, SceneSplit~\cite{lee2026jailbreaking}, SPARK~\cite{ying2025spark}, TFM~\cite{chen2026two} and BSB~\cite{peng2026between}. We first employ ASR@B~\cite{chao2024jailbreakbench}, commonly used in prior work, to measure the baseline attack success rate within a given query budget. Building on this, and in light of the temporal characteristics of T2V attacks and query-constrained scenarios, we further propose three evaluation metrics: TP-ASR@B requires the generated results to simultaneously satisfy the generation of dangerous content, preservation of the original intent, and temporal continuity, and is used to measure the strict attack success rate within a given budget; AUC-TP aggregates TP-ASR across different query budgets to characterise the overall attack performance throughout the entire finite query process; AvgQ measures the average number of target queries required to achieve a strict attack success for the first time; a lower value indicates that the method can achieve a successful attack with fewer queries. Under a fixed query budget, TempQ-Jail achieves TP-ASR@5 and TP-ASR@10 of 48.9\% and 65.4\%, respectively, representing improvements of 4.6 and 4.0 percentage points over the strongest baseline under the corresponding budgets. Across the complete query trajectory, TempQ-Jail also achieves the highest AUC-TP (0.469) and the lowest AvgQ (6.3). Compared to the best baseline, it increases AUC-TP from 0.425 to 0.469 and reduces AvgQ from 6.8 to 6.3, indicating that it not only improves the success rate of strict attacks within a limited budget but also enables valid attack candidates to enter the actually accessible query sequence at an earlier stage. Overall, TempQ-Jail transforms large-scale heterogeneous jailbreak candidates into attack sequences that can be efficiently utilised within a limited target query budget, providing a more effective black-box attack method for high-cost T2V systems.

The main contributions of this paper are as follows:

\begin{itemize}
    \item We model black-box jailbreaking in secure T2V systems as a budget-constrained query allocation problem, emphasising that, under a finite budget, the effectiveness of an attack depends not only on the existence of valid candidates but also on whether they can be accessed in a timely manner within the query trajectory.
    
    \item We propose TempQ-Jail, comprising three core modules: heterogeneous candidate construction, end-to-end attack value estimation, and budget-aware candidate ranking. Based on a comprehensive evaluation of candidates' attack potential, it prioritises the allocation of high-value candidates to limited target queries.
    
    \item Under a unified guarded T2V evaluation, TempQ-Jail achieves the best results across all evaluation metrics amongst the seven compared methods, and consistently outperforms the other methods under limited target query budgets.
\end{itemize}


\section{Related Work}
\label{sec:related_work}

\subsection{Text-to-Video Generation}
\label{sec:t2v_generation}

Diffusion models were first carried over from images to video by modelling the temporal axis jointly with the spatial ones~\cite{ho2022video}; cascaded pixel-space pipelines then showed that high-definition video could be synthesised directly from text~\cite{ho2022imagen}, and paired text-video supervision was shown to be dispensable by transferring image priors into the video domain~\cite{singer2022make}. To avoid the cost of operating in pixel space, subsequent work moved generation into a learned latent space and inflated pretrained image backbones with temporal layers~\cite{blattmann2023align,luo2023videofusion}, whilst other approaches obtained video generation from text-to-image models with little or no video-specific training~\cite{khachatryan2023text2video,guo2023animatediff}. More recently, transformer-based latent diffusion~\cite{ma2024latte}, space-time architectures that produce the whole temporal extent in a single pass~\cite{bar2024lumiere}, and increasingly capable open-source systems~\cite{wang2023modelscope,zheng2024open} have markedly improved motion coherence and the duration of the generated clips.

This progression is what turns the temporal axis into a security concern in its own right: a model able to sustain a coherent event across many frames is equally able to sustain a harmful one. It is precisely this property that our attack success criterion targets, and that distinguishes the T2V setting from the static image case.

\subsection{Jailbreak Attacks on T2V Systems}
\label{sec:t2v_jailbreak_attacks}

Jailbreak research on guarded generative models began in the text-to-image setting, and the mechanisms developed there are the direct antecedents of current T2V attacks~\cite{li2026sok}. Query-based search perturbs the tokens of a blocked prompt and uses the target's responses to steer the search, which establishes the number of queries as an explicit cost of attacking a guarded system~\cite{yang2024sneakyprompt,liang2022parallel,liang2022large,liang2025hard,wang2025black,wang2025no}; concept-inversion methods construct prompts that carry a removed concept in a form the filter no longer recognises~\cite{tsai2024ring}; and automated red teaming searches for problematic prompts that expose the gap between a model's safety mechanism and its generative capability~\cite{chin2023prompting4debugging, guo2026wmattack,xu2026ctrlattack,wang2026text}. Later work widened the attack surface across modalities~\cite{yang2024mma,he2023sa,liu2025bridging,zhang2026visual,kong2025universal,kong2024patch,liu2023improving}, made adversarial prompt construction controllable~\cite{ma2025jailbreaking}, and showed that individually benign text and image components can combine into unsafe output~\cite{liu2025multimodal}, whilst systematic benchmarking has since placed these attacks on a common footing~\cite{jin2025jailbreakdiffbench,ying2026safebench,liu2025agentsafe,zhang2025bench2advlm,liu2025metadv}. These studies supply the candidate construction techniques that T2V attacks inherit, but they operate on a single static image, so neither temporal persistence nor the much higher per-query cost of video generation arises as a constraint.

As T2V systems have progressively introduced input security filtering and content moderation mechanisms, related research has begun to explore how to circumvent these security defences whilst retaining the original malicious intent~\cite{liu2026t2v,lee2026jailbreaking,ying2025spark,chen2026two,peng2026between}. Existing T2V jailbreak methods primarily focus on the construction and search for attack candidates. One class of methods reduces the explicitness of dangerous semantics through semantic rewriting, implicit expression or narrative contextualisation~\cite{liu2026t2v,ying2025spark}; another class utilises conceptual decomposition, breaking down complete dangerous events into objects, actions, scenes and their relationships, before reorganising them into more covert attack prompts, such as DACA~\cite{deng2023divide}. Furthermore, optimisation-based methods such as T2V-OptJail strike a balance between evading security filters and preserving attack intent by searching for candidate expressions~\cite{liu2026t2v}. Whilst such methods enhance the stealthiness of the candidates, there remains significant uncertainty as to whether the candidates can be used to generate videos that align with the original intent and feature sustained dangerous events.

Subsequent work has further expanded the candidate space by focusing on aspects such as scene organisation, prompt reconstruction and search strategies. SceneSplit reduces the explicit exposure of dangerous intent by splitting and reorganising scene semantics~\cite{lee2026jailbreaking}; SPARK and TFM explore different mechanisms for prompt reconstruction and attack search~\cite{ying2025spark,chen2026two}; whilst BSB focuses on constructing and searching for attack candidates around security decision boundaries~\cite{peng2026between}. These methods further enhance candidate generation capabilities, but their primary focus remains on obtaining more or more effective candidates, with less consideration given to access priorities amongst candidates; when candidates pass security filtering yet fail to generate dangerous content, deviate from the original intent, or lack temporal continuity, limited queries may still be consumed by low-value candidates.

Overall, existing T2V jailbreak methods primarily address how to generate valid candidates, whilst paying insufficient attention to which candidates should be prioritised under limited target queries. This paper therefore further investigates end-to-end value assessment and budget-aware ranking of heterogeneous candidates, enabling those with greater attack potential to be prioritised within the limited query trajectory.

\subsection{Safety Evaluation and Safeguarding for T2V Models}
\label{sec:t2v_safety_evaluation}

As T2V models rapidly improve in terms of generation quality, motion consistency and the ability to model complex events, the risk of them generating violent, pornographic, self-harm, illegal and other unsafe content has attracted increasing attention~\cite{miao2024t2vsafetybench,dai2024safesora}. Existing research has gradually expanded from traditional video quality evaluation~\cite{huang2024vbench,liu2024evalcrafter} to systematic assessments focused on generation safety~\cite{miao2024t2vsafetybench,dai2024safesora}. T2VSafetyBench constructs malicious prompts across multiple safety dimensions and systematically evaluates the safety risks of different T2V models~\cite{miao2024t2vsafetybench}; simultaneously, video generation possesses temporal safety characteristics distinct from those of static images, with certain hazardous semantics requiring cross-frame actions, event evolution and their combinations to be fully realised~\cite{miao2024t2vsafetybench,lee2026jailbreaking,chen2026two}. Subsequent research has further extended security evaluation to more complex conditional video generation scenarios; for example, SafeGen-Bench focuses on the risks of harmful video generation under combined image and text conditions, and further reveals the challenges faced by security measures in multimodal contexts~\cite{ma2026safegen}.

Corresponding to the aforementioned attack studies, existing work has also developed T2V security mechanisms at different stages, including the input, generation process and output~\cite{liang2026t2vshield,dai2024safesora}. Input-side methods primarily utilise prompt moderation, semantic detection or prompt sanitisation to identify and block potentially dangerous requests prior to generation~\cite{liang2026t2vshield}, following the guard models introduced for text-to-image systems~\cite{yang2024guardt2i}; generation-side methods reduce the probability of generating unsafe content through safety concept suppression, conditional intervention or safety guidance~\cite{dai2024safesora}, adapting techniques first developed for image diffusion, where unsafe directions are suppressed during sampling~\cite{schramowski2023safe} or the corresponding concepts are erased from the model weights~\cite{gandikota2023erasing}; whilst output-side methods further combine visual content with temporal information to detect potential risks in generated videos~\cite{liang2026t2vshield}. For example, T2VShield constructs T2V jailbreak defences across two stages: input prompt cleansing and video-side risk detection~\cite{liang2026t2vshield}, whilst other works have further explored security control mechanisms tailored to multimodal conditions and the generation process~\cite{ma2026safegen,wang2026runawayevil}. Collectively, these studies constitute the primary security framework for current guarded T2V systems and further illustrate that the evaluation of attack methods must not be limited to whether input filtering is bypassed, but must also take into account the generation of dangerous content, the maintenance of attack intent, and temporal continuity within the video~\cite{liang2025safemobile}.


\section{Problem Definition}
\label{sec:problem_definition}

This paper investigates query-constrained black-box jailbreaks in T2V systems with input security filtering mechanisms. Given an initial malicious intent, an attacker can construct multiple candidate prompts locally but is limited to a finite number of queries to the target system. Our objective is not to further increase the number of candidates, but rather to determine the access priority of candidates within a fixed candidate pool, ensuring that those truly capable of achieving end-to-end attack success enter the finite query trajectory as early as possible.

\subsection{Problem Setting and Threat Model}
\label{sec:threat_model}

Given an original malicious intent $x$, the attacker first constructs locally a candidate pool $\mathcal{C}(x)$ containing $N$ candidate prompts:
\begin{equation}
\mathcal{C}(x)=\{c_1,c_2,\ldots,c_N\}.
\end{equation}

The target system consists of an input security gate and a victim T2V model. For a candidate $c \in \mathcal{C}(x)$, let $G(c) \in \{0, 1\}$ denote whether it passes the input security check. When $G(c)=1$, the candidate is fed into the victim model $\mathcal{M}$ to generate a video $v$; when $G(c)=0$, the candidate is intercepted directly and no video is generated. Regardless of whether the candidate passes the security gate, each submission of a candidate to the target system counts as one target query.

We consider a standard black-box attack setting, where the attacker has no access to the internal parameters, gradients or training data of the victim model or the security filter, and can only observe the system's return results through target queries. The attacker may train a surrogate model offline using historical interaction data independent of the final testing intent; however, during the formal testing process, the candidate generation mechanism, the surrogate model and the ranking strategy remain fixed, and candidates are not regenerated nor models updated online based on feedback from the target system. Consequently, the attacker must determine, within a finite target query budget, which candidates from a pre-constructed pool should be prioritised for access.

\subsection{Success of End-to-End Attacks}
\label{sec:end_to_end_success}

Merely passing through the security gate is not sufficient to constitute a successful T2V jailbreak. For a video $v$ generated from candidate $c$, we further consider three video-side conditions: $R(v)\in\{0,1\}$ indicates whether the video produces the target hazardous visual content, $I(v,x) \in \{0,1\}$ indicates whether the generated result preserves the original harmful intent $x$, and $P(v) \in \{0,1\}$ indicates whether the dangerous event satisfies the temporal continuity requirement.

Based on the above conditions, we first define the basic attack success indicator variable $y_{\mathrm{ASR}}(c)$ for candidate $c$:
\begin{equation}
y_{\mathrm{ASR}}(c)=G(c)\,R(v)\,I(v,x).
\end{equation}

Furthermore, we define a result that simultaneously satisfies the security gate passage, the generation of dangerous visual content, the preservation of the original intent, and temporal continuity as TP-Success, whose strict attack success indicator $y_{\mathrm{TP}}(c)$ is
\begin{equation}
y_{\mathrm{TP}}(c)=G(c)\,R(v)\,I(v,x)\,P(v).
\end{equation}

Consequently, TP-Success corresponds to the complete attack chain
\begin{equation}
\begin{aligned}
\text{Pass}
&\rightarrow\text{Risk Realisation}\\
&\rightarrow\text{Intent Preservation}\\
&\rightarrow\text{Temporal Persistence},
\end{aligned}
\end{equation}
and $y_{\mathrm{TP}}(c)=1$ necessarily implies that $y_{\mathrm{ASR}}(c)=1$. This strict definition distinguishes between candidates that merely bypass security filters and those that genuinely induce the victim model to generate target hazardous videos.

\subsection{Query-Limited Candidate Allocation}
\label{sec:query_limited_allocation}

Given a fixed candidate pool, the key to a query-limited attack lies in determining the order in which candidates are accessed. Let $\pi(x)$ denote the complete candidate ordering corresponding to the original intent $x$, where $c_{(b)}$ denotes the $b$th candidate in the ordering:
\begin{equation}
\pi(x)=\bigl(c_{(1)},c_{(2)},\ldots,c_{(N)}\bigr).
\end{equation}

Given a target query budget $B$, the attacker can access at most the first $B$ candidates in this ranking order. Let $\pi_B(x)$ denote the prefix of accessible candidates under budget $B$; then
\begin{equation}
\pi_B(x)=\operatorname{Prefix}_B\!\left(\pi(x)\right)=\bigl(c_{(1)},\ldots,c_{(B)}\bigr).
\end{equation}

To indicate whether an original intent can achieve a strict attack success within budget $B$, we define the budget-level success variable $Y_{\mathrm{TP}}(x,B)$:
\begin{equation}
Y_{\mathrm{TP}}(x,B)=\mathbb{1}\!\left[\exists\, b\leq B:\,y_{\mathrm{TP}}\!\left(c_{(b)}\right)=1\right].
\end{equation}

Consequently, the objective of query-constrained candidate allocation is not merely to ensure the existence of valid attacks within the candidate pool, but to position these valid candidates as far forward in the ranking as possible, so that they are actually accessed earlier under different finite budgets. For the maximum query budget $B_{\max}$, we express the overall objective as
\begin{equation}
\max_{\pi}\ \mathbb{E}_{x}\left[\frac{1}{B_{\max}}\sum_{B=1}^{B_{\max}}Y_{\mathrm{TP}}(x,B)\right].
\end{equation}

This objective takes multiple query budgets into account; the higher a strictly successful candidate ranks, the more likely it is to be accessed under a greater number of finite budget settings. Based on this problem definition, the next section describes how TempQ-Jail achieves this objective through heterogeneous candidate construction, end-to-end attack value estimation, and budget-aware candidate ranking.


\section{TempQ-Jail}
\label{sec:tempq_jail}

\subsection{Overview}
\label{sec:overview}

\begin{figure*}[t]
  \centering
  \includegraphics[width=\linewidth]{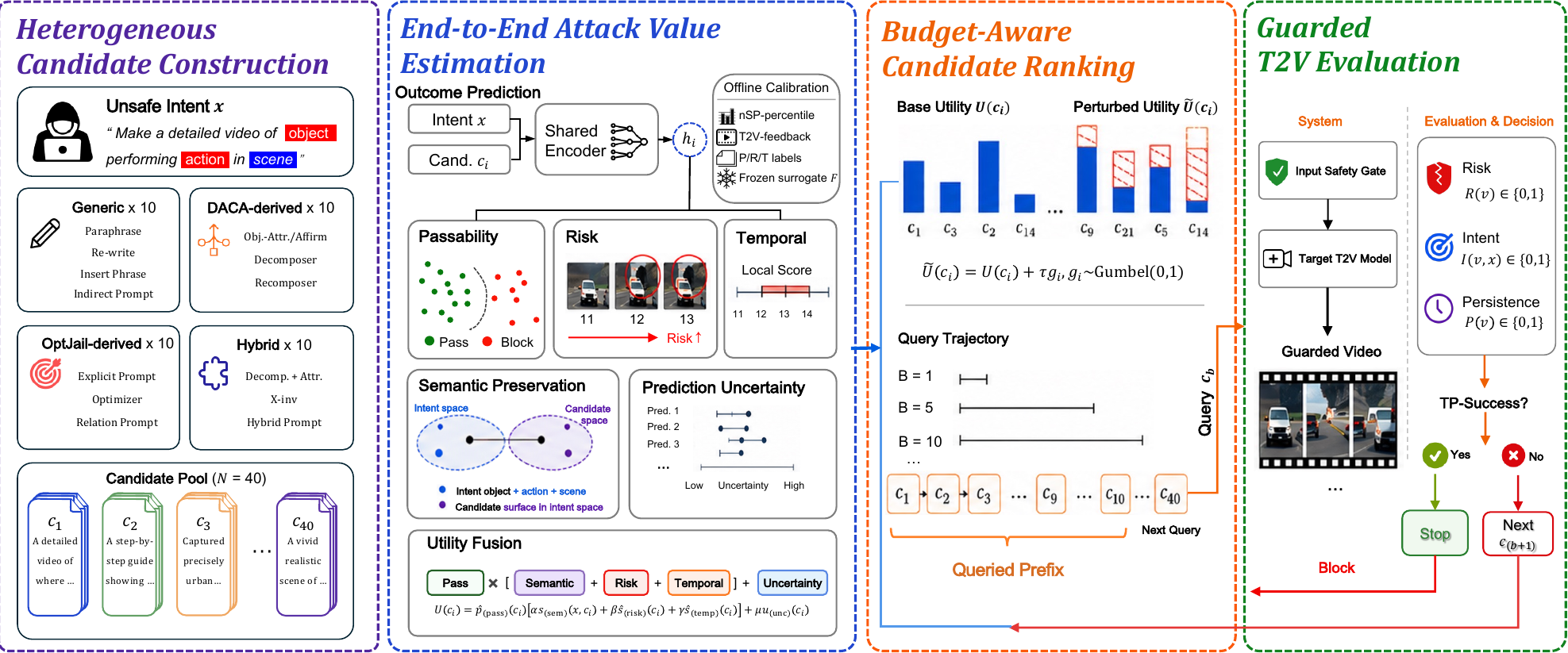}
  \caption{Overall workflow of TempQ-Jail. Heterogeneous Candidate Construction builds a
  fixed, diverse candidate pool for each unsafe intent; End-to-End Attack Value Estimation
  scores every candidate on gate passability, hazardous visual risk, temporal persistence,
  semantic preservation and prediction uncertainty; Budget-Aware Candidate Ranking turns
  these utilities into a single query priority sequence that is submitted to the guarded
  T2V system until a TP-Success occurs or the budget $B_{\max}$ is exhausted.}
  \label{fig:framework}
\end{figure*}

Based on the aforementioned query-constrained candidate allocation and ranking problem, we propose TempQ-Jail, whose core objective is to identify, amongst a large number of attack candidates, those most likely to achieve end-to-end attack success, and to prioritise the allocation of limited target queries to these candidates. Overall, TempQ-Jail comprises three core modules: Heterogeneous Candidate Construction, End-to-End Attack Value Estimation, and Budget-Aware Candidate Ranking. Firstly, the Heterogeneous Candidate Construction module draws upon a variety of attack mechanisms to establish a fixed and diverse pool of candidates for each original malicious intent, thereby enhancing the candidate space's ability to cover different attack manifestations; subsequently, the End-to-End Attack Value Estimation module comprehensively evaluates the attack potential of each candidate in terms of security gate passage, generation of dangerous visual content, preservation of original intent and temporal validity, and combines this with prediction uncertainty to derive a unified candidate value; finally, the Budget-Aware Candidate Ranking module generates a complete query priority sequence based on candidate value, enabling candidates with greater attack potential to enter the finite query trajectory earlier, thereby improving the attack success rate under a limited budget. Fig.~\ref{fig:framework} illustrates the overall workflow of TempQ-Jail.

\subsection{Heterogeneous Candidate Construction}
\label{sec:heterogeneous_candidate_construction}

Under a finite query budget, the success of an attack depends primarily on whether the candidate pool contains valid attack candidates. If the candidate space itself is under-covered, it is difficult to achieve a successful attack within a limited number of queries, even if subsequent evaluation and ranking are performed accurately. Existing T2V jailbreak methods typically rely on specific candidate generation mechanisms~\cite{liu2026t2v,lee2026jailbreaking,ying2025spark,chen2026two,peng2026between}, with different mechanisms placing varying emphasis on linguistic expression, semantic obfuscation and the organisation of temporal events. Therefore, to improve the likelihood of obtaining valid attack candidates within a limited number of queries, TempQ-Jail first integrates multiple complementary attack mechanisms to construct a fixed yet diverse heterogeneous candidate pool, thereby expanding the attack coverage capability of the candidate space.

For a given original malicious intent $x$, we construct four distinct classes of candidates: Generic, DACA-derived~\cite{deng2023divide}, T2V-OptJail-derived~\cite{liu2026t2v} and Hybrid candidate sets, denoted respectively as $\mathcal{C}_{\mathrm{gen}}(x)$, $\mathcal{C}_{\mathrm{daca}}(x)$, $\mathcal{C}_{\mathrm{opt}}(x)$ and $\mathcal{C}_{\mathrm{hyb}}(x)$. The complete candidate pool $\mathcal{C}(x)$ is defined as
\begin{equation}
\mathcal{C}(x)=\mathcal{C}_{\mathrm{gen}}(x)\cup\mathcal{C}_{\mathrm{daca}}(x)\cup\mathcal{C}_{\mathrm{opt}}(x)\cup\mathcal{C}_{\mathrm{hyb}}(x).
\end{equation}

As each of the four candidate categories contains the same number of prompts, the total size $N$ of the candidate pool is
\begin{equation}
\begin{aligned}
\left|\mathcal{C}_{\mathrm{gen}}(x)\right|
&=
\left|\mathcal{C}_{\mathrm{daca}}(x)\right|
=
\left|\mathcal{C}_{\mathrm{opt}}(x)\right|\\
&=
\left|\mathcal{C}_{\mathrm{hyb}}(x)\right|,\\
N&=\left|\mathcal{C}(x)\right|.
\end{aligned}
\end{equation}

The four categories of candidates expand the attack expression space from different perspectives. Generic candidates increase the diversity of linguistic expression through semantic rewriting, narrative context adjustment, synonymous expressions and changes in descriptive granularity; DACA-derived candidates draw on the concept of decomposition, breaking down complete hazardous events into objects, actions, environments and their relationships, before reorganising the candidate prompts using more dispersed and indirect language; T2V-OptJail-derived candidates draw on the joint consideration of filter evasion and attack intent in optimised jailbreak approaches to generate candidate prompts with varying degrees of explicitness and surface expressions; Hybrid candidates, meanwhile, further integrate semantic rewriting, conceptual decomposition and temporal event reconstruction to cover combinatorial attack forms that are difficult to express using a single attack mechanism.

The four categories of candidates have the same scale within the initial candidate pool; TempQ-Jail submits all candidates to the subsequent end-to-end attack value estimation module for unified evaluation, thereby determining which candidates are more worthy of using the limited number of target queries.

\subsection{End-to-End Attack Value Estimation}
\label{sec:end_to_end_attack_value}

Heterogeneous candidate constructs can expand the potential attack space, but the actual effectiveness of candidates remains difficult to predict in advance. Even if a candidate is able to bypass input security filtering, it may fail to induce the model to generate dangerous visual content, or the generated results may deviate from the original attack intent or lack sustained temporal behaviour. Therefore, to address the challenge of the difficulty in predicting the effectiveness of candidate attacks, TempQ-Jail has designed an end-to-end attack value estimation module. This module evaluates the attack potential of each candidate based on the complete attack chain, thereby identifying candidates that are more worthy of utilising the limited number of target queries.

Given an original malicious intent $x$ and its candidate prompts $c\in\mathcal{C}(x)$, we first use a text encoder $f_{\theta}$ to obtain the $d$-dimensional representation vector $\mathbf{h}_c\in\mathbb{R}^{d}$ for the candidate:
\begin{equation}
\mathbf{h}_c=f_{\theta}(x,c).
\end{equation}

Based on the candidate representation $\mathbf{h}_c$, the surrogate model utilises three prediction heads to estimate, respectively, the candidate's probability of passing the safety gate $\hat p_{\mathrm{pass}}(c)$, the risk visual generation score $\hat s_{\mathrm{risk}}(c)$ and the temporal validity score $\hat s_{\mathrm{temp}}(c)$:
\begin{equation}
\begin{aligned}
\hat p_{\mathrm{pass}}(c)&=g_{\mathrm{pass}}(\mathbf{h}_c),\\
\hat s_{\mathrm{risk}}(c)&=g_{\mathrm{risk}}(\mathbf{h}_c),\\
\hat s_{\mathrm{temp}}(c)&=g_{\mathrm{temp}}(\mathbf{h}_c).
\end{aligned}
\end{equation}

Here, $\hat p_{\mathrm{pass}}(c)$ measures the likelihood that the candidate will pass through the safety gate and enter the video generation stage; $\hat s_{\mathrm{risk}}(c)$ measures the candidate's potential to induce harmful visual content; and $\hat s_{\mathrm{temp}}(c)$ measures the potential for harmful events to persist over the temporal dimension of the video. In addition, we use the scalar $s_{\mathrm{sem}}(x,c)$ to measure the degree of semantic preservation between candidate $c$ and the original malicious intent $x$, thereby preventing the candidate from deviating from the original attack target whilst evading the filter.

Given the uncertainty in the surrogate model's predictions for different candidates, we employ multiple ensemble members with independent initialisations and utilise the discrepancies in their predictions regarding hazardous visual generation and temporal validity to estimate the uncertainty of the candidate. Let there be $M$ members in the ensemble; the $m$th member's predictions for the candidate $c$ regarding visual risk and temporal validity are denoted as $\hat s_{\mathrm{risk}}^{(m)}(c)$ and $\hat s_{\mathrm{temp}}^{(m)}(c)$ respectively. The uncertainty score $u_{\mathrm{unc}}(c)$ for the candidate is then defined as
\begin{equation}
u_{\mathrm{unc}}(c)
=
\frac{1}{2}
\left(
\operatorname{Var}_{m=1}^{M}
\!\left[\hat s_{\mathrm{risk}}^{(m)}(c)\right]
+
\operatorname{Var}_{m=1}^{M}
\!\left[\hat s_{\mathrm{temp}}^{(m)}(c)\right]
\right).
\end{equation}

A larger value of $u_{\mathrm{unc}}(c)$ indicates greater predictive divergence amongst the ensemble members regarding the attack effectiveness of the current candidate. This signal is used to preserve a limited capacity for exploration during candidate evaluation.

Finally, we integrate the above signals into an end-to-end attack utility $U(c)$ for candidate $c$. Here, $\alpha$, $\beta$ and $\gamma$ control the contributions of the semantic preservation, dangerous visual generation and temporal validity signals respectively, whilst $\mu$ controls the contribution of prediction uncertainty:
\begin{equation}
\begin{aligned}
U(c)
={}&
\hat p_{\mathrm{pass}}(c)
\bigl[
\alpha s_{\mathrm{sem}}(x,c)
+
\beta\hat s_{\mathrm{risk}}(c)\\
&\qquad\quad
+
\gamma\hat s_{\mathrm{temp}}(c)
\bigr]
+
\mu u_{\mathrm{unc}}(c).
\end{aligned}
\end{equation}

Here, the probability of passing the security gate, $\hat p_{\mathrm{pass}}(c)$, acts on the primary attack value term, causing candidates that are difficult to pass through input filtering to receive a lower overall utility; semantic preservation, hazardous visual generation and temporal validity collectively measure a candidate's potential to realise a complete attack chain; higher uncertainty indicates significant divergence in predictions amongst different ensemble members regarding the effectiveness of candidate attacks. We incorporate this into the candidate utility with a smaller weighting, reserving a certain ranking priority for candidates with unstable predictions, thereby reducing the risk that potentially valid candidates are prematurely ranked lower due to estimation errors in the surrogate model.

Through the above end-to-end evaluation, $U(c)$ comprehensively reflects a candidate's attack potential, from passing through the input security gate to generating dangerous content that aligns with the original attack intent and exhibits temporal persistence. The next module further utilises this candidate value to generate a unified query priority sequence, enabling high-value candidates to be accessed prioritarily within a limited target query budget.

\subsection{Budget-Aware Candidate Ranking}
\label{sec:budget_aware_candidate_ranking}

End-to-end attack value estimation can determine the potential impact of different candidates; however, in scenarios where the number of target queries is limited, the order in which candidates are accessed also directly affects the final attack outcome. Even if the candidate pool contains prompts capable of achieving a successful attack, if their position in the query sequence is too late, they may not be actually accessed before the budget is exhausted. Therefore, to address the challenge of a limited query budget, TempQ-Jail further converts the candidate-level end-to-end attack utility obtained in the previous section into a unified query priority sequence, ensuring that high-value candidates are placed as far forward as possible within the limited budget.

For a candidate $c \in \mathcal{C}(x)$, the end-to-end attack utility $U(c)$ has already been obtained in the previous section. When multiple candidates have similar predicted utilities, their relative order is susceptible to subtle prediction errors in the surrogate model. To prevent the query sequence from being entirely determined by these minor differences, we introduce low-temperature Gumbel perturbation to the candidate utility. Let $g_c$ denote the Gumbel random variable corresponding to candidate $c$, and let $\tau$ be the temperature coefficient controlling the strength of the perturbation; the perturbed candidate utility is denoted as $\widetilde U(c)$:
\begin{equation}
\widetilde U(c)=U(c)+\tau g_c,\qquad
g_c\sim\mathrm{Gumbel}(0,1).
\end{equation}

Here, the end-to-end attack utility $U(c)$ continues to determine the primary priority of the candidate, whilst the Gumbel perturbation introduces only a limited change in order amongst candidates with similar utilities, thereby reducing the impact of the ranking's over-reliance on minute prediction differences in the surrogate model.

Subsequently, TempQ-Jail ranks all candidates in the candidate pool in descending order based on $\widetilde U(c)$. Let $\pi(x)$ denote the complete candidate ranking corresponding to the original malicious intent $x$, and let $c_{(b)}$ denote the candidate ranked $b$th in priority; then
\begin{equation}
\begin{aligned}
&\pi(x)=\bigl(c_{(1)},c_{(2)},\ldots,c_{(N)}\bigr),\\
&\widetilde U(c_{(1)})
\geq
\widetilde U(c_{(2)})
\geq
\cdots
\geq
\widetilde U(c_{(N)}).
\end{aligned}
\end{equation}

In query-constrained scenarios, the budget does not alter the relative ordering of the candidates, but rather determines how many candidates in the complete sequence can actually be accessed. Specifically, given a target query budget $B$, let $\pi_B(x)$ denote the accessible candidate prefixes within the budget; then
\begin{equation}
\pi_B(x)
=
\operatorname{Prefix}_{B}\!\left(\pi(x)\right)
=
\bigl(c_{(1)},c_{(2)},\ldots,c_{(B)}\bigr).
\end{equation}

All query budgets share the same complete ordering. For any two budgets $B_1$ and $B_2$, when $B_1 < B_2$, the candidate sequence corresponding to the smaller budget is always a prefix of the sequence for the larger budget:
\begin{equation}
\pi_{B_1}(x)
=
\operatorname{Prefix}_{B_1}\!\left(\pi_{B_2}(x)\right),
\qquad
B_1<B_2.
\end{equation}

This nested prefix design ensures that different query budgets correspond to the same continuous query trajectory, rather than regenerating candidates or re-executing the ranking for each budget. Consequently, a higher candidate priority implies that the candidate can actually be accessed under more constrained budget settings; this is the key mechanism by which TempQ-Jail links candidate value to a finite query budget.

During an actual attack, TempQ-Jail submits candidates to the guarded T2V system in ranked order, starting from $c_{(1)}$. If the current candidate satisfies the strict attack success conditions defined earlier, the current query process is terminated immediately; otherwise, access proceeds to the next candidate $c_{(b+1)}$, until the attack succeeds or the given query budget $B$ is reached. Consequently, $\pi_B(x)$ represents the maximum candidate prefix that can be accessed within the budget, whilst the actual query trajectory may be shorter than this prefix due to early success.

Through the above design, TempQ-Jail translates the end-to-end attack value at the candidate level into an actual access sequence under a finite budget. Its core mechanism does not involve repeatedly optimising the ordering for different budgets, but rather prioritises high-value candidates at the beginning, ensuring that potentially effective attacks are queried first within the shorter accessible prefix. This enhances the utilisation efficiency of the limited target queries and the attack success rate within the budget.

\subsection{Overall Optimisation Pipeline}
\label{sec:overall_pipeline}

\begin{small}
\noindent\textbf{Algorithm 1: TempQ-Jail}

\noindent\textbf{Input:} Original malicious intent $x$, maximum query budget $B_{\max}$, frozen surrogate model $\mathcal{F}$

\noindent\textbf{Output:} Query trajectory and attack results

\begin{enumerate}
    \item Construct a heterogeneous candidate pool $\mathcal{C}(x)$ from four candidate generation mechanisms;
    
    \item For each candidate $c\in\mathcal{C}(x)$:
    \begin{itemize}
        \item Compute the candidate representation and semantic retention score;
        \item Predict passability, visual risk and temporal effectiveness;
        \item Compute the ensemble prediction uncertainty;
        \item Compute $U(c)$ based on the joint attack utility;
        \item Sample Gumbel noise and compute $\widetilde U(c)$;
    \end{itemize}
    
    \item Generate a complete candidate ranking $\pi(x)$ in descending order based on $\widetilde U(c)$;
    
    \item Query the target T2V system sequentially, starting from $c_{(1)}$;
    
    \item If the current candidate achieves a strict attack success, terminate;
    
    \item Otherwise, continue querying $c_{(b+1)}$ until $B_{\max}$ is reached.
\end{enumerate}
\end{small}

Algorithm 1 summarises the overall optimisation workflow of TempQ-Jail. Given a source malicious intent $x$ and a maximum target query budget $B_{\max}$, TempQ-Jail first constructs a fixed heterogeneous candidate pool $\mathcal{C}(x)$ based on the four candidate generation mechanisms introduced in Sec.~\ref{sec:heterogeneous_candidate_construction}. For each candidate $c\in\mathcal{C}(x)$, the frozen surrogate model estimates its probability of passing through the security gate, its potential for generating dangerous visuals, and its temporal validity, and calculates the end-to-end attack utility $U(c)$ by combining the degree of semantic preservation with ensemble prediction uncertainty. Subsequently, TempQ-Jail generates a unified candidate priority sequence $\pi(x)$ based on the budget-aware ranking strategy described in Sec.~\ref{sec:budget_aware_candidate_ranking}. During the target system interaction phase, candidates are submitted to the guarded T2V system in the order of this ranking until either a strict attack succeeds for the first time or the maximum query budget $B_{\max}$ is exhausted. Throughout the testing process, the candidate generation mechanism, surrogate model, utility function and ranking rules remain fixed; no candidate regeneration, surrogate model updates or online re-ranking based on target feedback are performed.

In the specific implementation, we construct $N=40$ candidates for each original malicious intent, comprising 10 candidates each from the four categories: Generic, DACA-derived, T2V-OptJail-derived and Hybrid. The candidate texts are encoded into 384-dimensional representations using SentenceBERT \texttt{all-MiniLM-L6-v2}. The surrogate model consists of five independently initialised MLPs and is trained and calibrated offline using 450 warm-up candidate-query examples drawn from outside the final test set of harmful intents. The surrogate model was trained for 10 epochs using the Adam optimiser, with a learning rate set to $10^{-3}$ and a batch size of 32.

The weights in the joint attack utility are set to $\alpha=0.30$, $\beta=0.25$, $\gamma=0.25$ and $\mu=0.10$, whilst the Gumbel temperature is set to $\tau=0.1$. All surrogate model parameters and hyperparameters are fixed prior to the start of the final test.


\section{Experiment}
\label{sec:experiment}

\subsection{Experimental Setup}
\label{sec:experimental_setup}

\paragraph{Victim Model.}
We use CogVideoX-5B~\cite{yang2025cogvideox} as the text-to-video (T2V) victim model and apply a uniform video generation configuration to all methods, with a resolution of $480\times480$, generating 17 video frames (approximately 8 fps), using 20 denoising steps and a guidance scale of 6.0, and fixing the generation seed to 1234 to ensure that the generated results for the same candidate prompts remain consistent across different methods and during repeated evaluations. To construct a unified guarded T2V system, we further deployed a text security filter based on GPT-4O-Mini at the input stage of the victim model. This filter acts as a surface-level text moderation gate, outputting either ``ALLOW'' or ``BLOCK'' based on the textual content of the candidate prompts, to intercept prompts that directly and explicitly describe unsafe content; only prompts that pass moderation are fed into CogVideoX-5B for video generation. Consequently, attackers must not only circumvent the input-side security filter but also ensure that prompts passing the review genuinely induce the generator to produce dangerous content that aligns with the original attack intent and exhibits temporal continuity. In this guarded T2V system, we define the submission of a candidate prompt to the target system as a single target query. Each candidate first undergoes input security review; if the result is ``ALLOW'', video generation proceeds and a subsequent attack success determination is carried out; if the result is ``BLOCK'', video generation is not executed, but the submission is still counted towards the target query budget. For each original malicious intent, we uniformly set the maximum target query budget to $B_{\max}=10$. All comparison methods utilise exactly the same victim model, video generation configuration, input security filter, video evaluator, attack success criteria and target query budget; the only differences between methods lie in the construction of candidate prompts and their query order, thereby ensuring a fair comparison of different attack strategies under a unified query-constrained protocol.

\paragraph{Dataset.}
We constructed a unified test set based on the malicious prompt data from T2VSafetyBench~\cite{miao2024t2vsafetybench}. This benchmark provides a total of 90 unsafe intents, categorised into different security risk classes according to the original annotations. As different attack methods vary in their ability to generate candidate prompts, some methods are unable to provide a sufficient number of candidates for all original intents to meet the requirements of the unified query protocol. To ensure that all methods are compared under exactly the same test intents and query conditions, we retained those original intents for which all compared methods were able to generate at least $B_{\max}$ viable attack candidates, and defined these samples that jointly satisfy the evaluation criteria as common viable intents. Following screening, we ultimately identified 70 common viable intents covering six categories of security risks: Violence, Gore, Self-harm, Pornography, Illegal Activity and Disturbing Content, and used these as a unified test set shared by all methods.

\paragraph{Attack Methods.}
We compare TempQ-Jail with six representative T2V jailbreak methods, including DACA~\cite{deng2023divide}, T2V-OptJail~\cite{liu2026t2v}, SceneSplit~\cite{lee2026jailbreaking}, SPARK~\cite{ying2025spark}, TFM~\cite{chen2026two} and BSB~\cite{peng2026between}. DACA constructs implicit attack prompts through conceptual decomposition and recombination; T2V-OptJail balances the evasion of security filters with the preservation of attack semantics by optimising candidate expressions; SceneSplit reduces the explicitness of dangerous semantics through scene splitting; SPARK and TFM represent distinct prompt reconstruction and attack search strategies, respectively; whilst BSB generates attack prompts closer to the security decision boundary by constructing boundary candidates.

\paragraph{Evaluation Metrics.}
Let the test set contain $n$ original malicious intents $\{x_i\}_{i=1}^{n}$, and let $c_{i,(b)}$ denote the $b$th candidate in the query sequence of intent $x_i$. Following the definition in Sec.~\ref{sec:problem_definition}, $y_{\mathrm{ASR}}(c_{i,(b)})=1$ indicates that this candidate passes through the input security gate and generates dangerous visual content consistent with the original malicious intent; furthermore, $y_{\mathrm{TP}}(c_{i,(b)})=1$ indicates that this candidate also satisfies temporal persistence; we refer to this stricter attack outcome as TP-Success. We first adopt ASR@$B$~\cite{chao2024jailbreakbench}, commonly used in existing jailbreak evaluations, to measure the proportion of attempts where at least one basic attack succeeds within the first $B$ target queries:
\begin{equation}
\mathrm{ASR@}B=
\frac{1}{n}\sum_{i=1}^{n}
\mathbb{1}\!\left[
\max_{1\leq b\leq B}
y_{\mathrm{ASR}}(c_{i,(b)})=1
\right].
\end{equation}

Building on this, we have further designed three evaluation metrics for query-constrained T2V jailbreaks. TP-ASR@$B$ measures the proportion of attempts in which at least one TP-Success is achieved within the first $B$ target queries:
\begin{equation}
\mathrm{TP\text{-}ASR@}B=
\frac{1}{n}\sum_{i=1}^{n}
\mathbb{1}\!\left[
\max_{1\leq b\leq B}
y_{\mathrm{TP}}(c_{i,(b)})=1
\right].
\end{equation}

AUC-TP averages the TP-ASR values from 1 up to the maximum query budget $B_{\max}$ to measure the overall strict attack performance across the entire finite query trajectory:
\begin{equation}
\mathrm{AUC\text{-}TP}
=
\frac{1}{B_{\max}}
\sum_{B=1}^{B_{\max}}
\mathrm{TP\text{-}ASR@}B.
\end{equation}

Finally, for an intent $x_i$, let $q_i^{*}$ denote the query position at which $y_{\mathrm{TP}}(c_{i,(b)})=1$ is first satisfied; if a successful strict attack is never achieved within $B_{\max}$ queries, then let $q_i^{*}=B_{\max}+1$. Accordingly, AvgQ is defined as
\begin{equation}
\mathrm{AvgQ}
=
\frac{1}{n}
\sum_{i=1}^{n}
q_i^{*},
\end{equation}
which represents the average number of target queries required to achieve the first successful strict attack; a lower value indicates that a successful attack can occur earlier in the query sequence.

\subsection{Overall Attack Performance}
\label{sec:overall_attack_performance}

We compared TempQ-Jail with six representative T2V jailbreak methods---DACA, T2V-OptJail, SceneSplit, SPARK, TFM and BSB---on CogVideoX-5B, using 70 common viable intents derived from T2VSafetyBench as a unified test set. All methods utilised the same victim model, safety filter, video generation configuration, success criteria and target query budget. We report ASR and TP-ASR under two representative budgets, $B=5$ and $B=10$, and further evaluate attack performance across the entire finite query trajectory using AUC-TP and AvgQ. The specific results are shown in Table~\ref{tab:overall_attack_performance}.

\begin{table*}[t]
  \centering
  \renewcommand{\arraystretch}{1.15}
  \setlength{\tabcolsep}{4mm}
  \begin{tabular}{lcccccc}
    \hline
    Method               & ASR@5 $\uparrow$ & TP-ASR@5 $\uparrow$ & ASR@10 $\uparrow$ & TP-ASR@10 $\uparrow$ & AUC-TP $\uparrow$ & AvgQ $\downarrow$ \\
    \hline
    DACA                 & 55.4             & 43.4                & 72.9              & 61.4                 & 0.419             & 6.8               \\
    T2V-OptJail          & 57.1             & 44.3                & 77.1              & 57.1                 & 0.425             & 6.8               \\
    SceneSplit           & 18.3             & 12.9                & 35.7              & 24.3                 & 0.143             & 9.6               \\
    SPARK                & 18.6             & 17.7                & 25.7              & 21.4                 & 0.165             & 9.4               \\
    TFM                  & 19.1             & 13.1                & 27.1              & 20.0                 & 0.129             & 9.7               \\
    BSB                  & 34.3             & 26.6                & 47.1              & 38.6                 & 0.261             & 8.4               \\
    \textbf{TempQ-Jail}  & \textbf{63.7}    & \textbf{48.9}       & \textbf{79.7}     & \textbf{65.4}        & \textbf{0.469}    & \textbf{6.3}      \\
    \hline
  \end{tabular}
  \caption{Overall attack performance comparison under different target query budgets.}
  \label{tab:overall_attack_performance}
\end{table*}

From Table~\ref{tab:overall_attack_performance}, we can draw the following conclusions: (1) TempQ-Jail achieves the highest strict attack success rate under a finite query budget. For example, when $B=5$, TempQ-Jail's TP-ASR@5 reaches 48.9\%, representing an improvement of 4.6 percentage points compared to the 44.3\% achieved by the strongest baseline, T2V-OptJail; when $B=10$, its TP-ASR@10 reaches 65.4\%, an improvement of 4.0 percentage points over the strongest baseline, DACA, which achieves 61.4\%. (2) TempQ-Jail also maintains a lead in terms of baseline attack success rate, indicating that its advantages are evident in both general attack success and more stringent temporal attack success. For example, its ASR@5 and ASR@10 reach 63.7\% and 79.7\% respectively, both exceeding all compared methods, whilst the corresponding strongest baselines are 57.1\% and 77.1\%. (3) TempQ-Jail enables strict successful candidates to enter the finite query trajectory earlier, thereby making more efficient use of the target query budget. For instance, its AUC-TP reaches a maximum of 0.469, exceeding the strongest baseline's 0.425, whilst AvgQ decreases from the baseline's best of 6.8 to 6.3, indicating that valid attacks can generally be accessed at earlier query positions.

\subsection{Query Efficiency under Limited Budgets}
\label{sec:query_efficiency}

\begin{figure}[t]
  \centering
  \includegraphics[width=\linewidth]{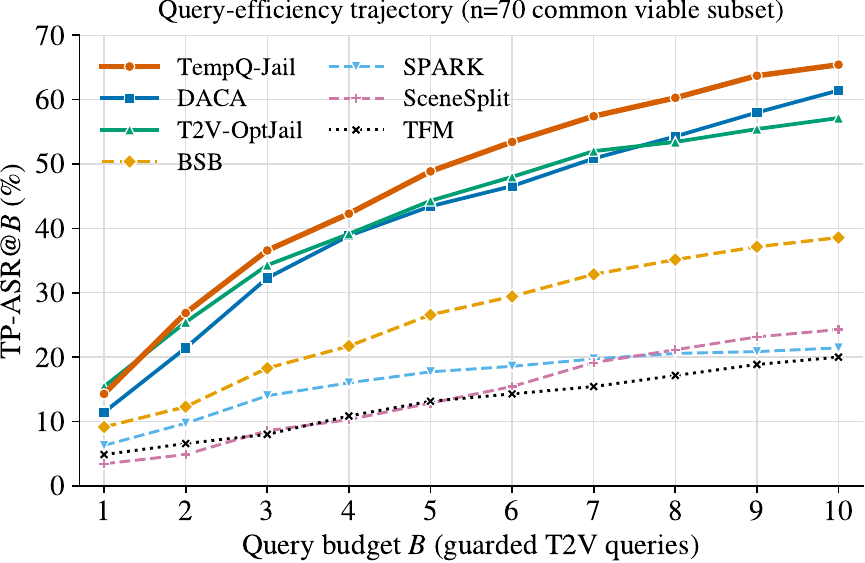}
  \caption{TP-ASR@$B$ under different target query budgets on the 70 common viable intents.
  All budgets share one nested query trajectory, so a higher curve at small $B$ means
  effective candidates are placed earlier in the query sequence.}
  \label{fig:query_efficiency}
\end{figure}

To further analyse the efficiency with which different attack methods utilise a limited target query budget, we examine the variation in TP-ASR@$B$ as the query budget is gradually increased from $B=1$ to $B_{\max}$, and compare the complete query trajectories of TempQ-Jail with those of six baseline methods in Fig.~\ref{fig:query_efficiency}. For each method and each original malicious intent, we first identify a fixed sequence of candidate queries; different query budgets correspond only to prefixes of varying lengths within that sequence. Consequently, an increase in the budget entails continuing to access subsequent candidates within the existing query trajectory, without regenerating candidates or reordering them to accommodate the new budget. Through this nested prefix evaluation, the trajectory of TP-ASR@$B$ as the budget varies directly reflects the position of strictly successful candidates within the query sequence; the earlier an effective candidate appears, the higher the strictly successful attack rate the method can achieve with a smaller query budget.

From Fig.~\ref{fig:query_efficiency}, we can draw the following conclusions: (1) The advantage of TempQ-Jail is primarily evident in the early stages of the query trajectory, enabling it to convert new queries into successful strict attacks more rapidly. At $B=1$, T2V-OptJail performs slightly better than TempQ-Jail; however, once the budget increases to $B=2$, TempQ-Jail achieves the highest TP-ASR and maintains its lead across the subsequent budget ranges. This indicates that TempQ-Jail does not rely on a large query budget to gradually ``stumble upon'' valid candidates, but rather is able to concentrate high-value candidates at the front of the query sequence at an early stage. (2) The relative advantages of different baseline methods vary significantly with changes in query budget, whereas TempQ-Jail demonstrates a more stable advantage across budget variations. For example, T2V-OptJail outperforms DACA overall in the low-budget range, but as the budget increases, DACA's performance curve continues to close the gap and eventually surpasses T2V-OptJail at higher budgets, suggesting that effective attacks generated by different candidate generation strategies may be distributed at different positions within the query sequence; in contrast, TempQ-Jail maintains a higher TP-ASR from smaller budgets onwards, suggesting that a unified candidate value ranking reduces the dependence of attack effectiveness on specific budget choices. (3) The improvement achieved by TempQ-Jail does not stem from any single budget point, but manifests as a sustained upward shift across the entire query trajectory. As $B$ increases from 2 to 10, its TP-ASR grows steadily and remains consistently higher than that of other methods, indicating that the increased number of query opportunities allows sustained access to candidates with higher attack value, rather than a rapid loss of benefit after the first few candidates. This result further validates the necessity of placing high-value candidates at the front of a unified query sequence.

\subsection{Discussion}
\label{sec:discussion}

\paragraph{Candidate Allocation.}
To analyse how TempQ-Jail reorganises candidates within a finite query prefix, we compare the source distributions of the initial candidate pool with those of the ranked Top-5 candidates. The initial candidate pool consists of four categories---Generic, DACA-derived, T2V-OptJail-derived and Hybrid---each accounting for 25\%; as actual queries may terminate prematurely upon success, the statistics here refer to the proportion of Top-5 ranked candidates rather than the proportion of queries actually executed.

\begin{table}[t]
  \centering
  \small
  \renewcommand{\arraystretch}{1.15}
  \setlength{\tabcolsep}{3mm}
  \begin{tabular}{lcc}
    \hline
    \multirow{2}{*}{Candidate Source} & Initial Pool & Top-5 Ranked \\
                                      & (\%)         & (\%)         \\
    \hline
    Generic                           & 25           & 8            \\
    DACA-derived                      & 25           & 34           \\
    T2V-OptJail-derived               & 25           & 32           \\
    Hybrid                            & 25           & 26           \\
    \hline
  \end{tabular}
  \caption{Candidate source distributions in the initial pool and the Top-5 ranked candidates.}
  \label{tab:candidate_allocation}
\end{table}

Two key observations can be drawn from Table~\ref{tab:candidate_allocation}: (1) Ranking significantly alters the composition of candidates in finite query prefixes. The proportions of DACA-derived and T2V-OptJail-derived candidates in the Top-5 increased to 34\% and 32\% respectively, whilst Generic candidates fell to 8\%, indicating that end-to-end attack value estimation can prioritise candidates with greater potential. (2) Query priority is determined by candidate-level value rather than being pre-specified by the candidate's source. The initial proportions of the four candidate categories were identical, but after unified ranking, they exhibited markedly different proportions at the front of the list, reflecting TempQ-Jail's adaptive priority allocation for heterogeneous candidates.

\begin{figure}[t]
  \centering
  \includegraphics[width=\linewidth]{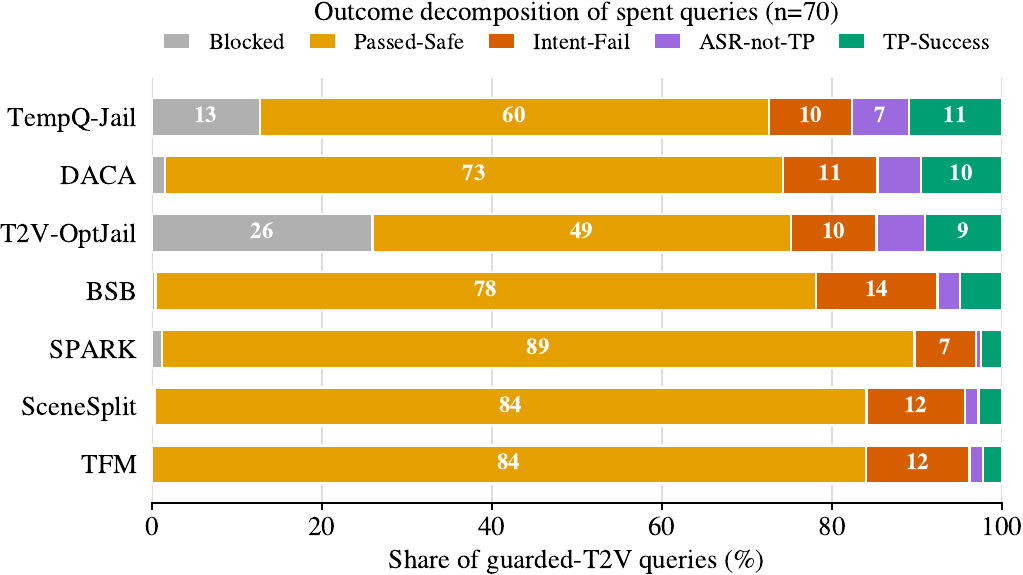}
  \caption{Failure attribution of the target queries actually executed by each attack method.
  Every query is assigned to exactly one of five outcomes, and querying stops after the first
  TP-Success; the TP-Success shares are therefore computed per query and are not directly
  comparable with the per-intent TP-ASR@$B$ values in Table~\ref{tab:overall_attack_performance}.}
  \label{fig:failure_attribution}
\end{figure}

\paragraph{Failure Attribution.}
To pinpoint where different attack methods fail within the complete attack chain, we classify all actually executed target queries into five mutually exclusive outcomes: ``Blocked'' indicates failure to pass the input security gate; ``Passed-Safe'' indicates passing the review but failing to generate the target hazardous content; ``Intent-Fail'' indicates generating hazardous content but deviating from the original attack intent; ``ASR-not-TP'' indicates achieving basic attack success but failing to satisfy temporal persistence; ``TP-Success'' indicates that the security gate was passed, dangerous visual content was generated, the intent was maintained, and temporal persistence was achieved. Fig.~\ref{fig:failure_attribution} shows the distribution of actually executed queries across these five categories, stopping after the first instance of ``TP-Success''; consequently, the ``TP-Success'' figures here are calculated per query and cannot be directly compared with the ``TP-ASR@$B$'' values in Table~\ref{tab:overall_attack_performance}, which are calculated per intent.

Two key observations can be drawn from Fig.~\ref{fig:failure_attribution}: (1) The primary failures of most methods occur after passing through the input safety gate. The ``Passed-Safe'' rates for DACA, SceneSplit, SPARK and TFM stand at 72.7\%, 83.7\%, 88.5\% and 84.0\% respectively, indicating that bypassing input filtering does not necessarily mean that valid, dangerous videos can be generated subsequently. (2) TempQ-Jail is able to convert a greater number of actual queries into complete attack successes. For example, whilst TFM has a ``Blocked'' rate of 0.0\%, its TP-Success rate is only 2.2\%; by contrast, whilst TempQ-Jail had 12.7\% of queries blocked, its TP-Success rate still reached a high of 11.0\%, indicating that end-to-end attack effectiveness depends on whether candidates can complete the full attack chain, including hazard generation, intent preservation, and temporal continuity, rather than merely passing through the input safety gate.

\paragraph{Ablation Study.}
To validate the effectiveness of TempQ-Jail's three core modules, we conducted ablation experiments using a step-by-step approach, starting with the baseline candidate strategy and sequentially incorporating Heterogeneous Candidate Construction (HCC), End-to-End Attack Value Estimation (EAVE) and Budget-Aware Candidate Ranking (BACR). All variants utilised the same victim model, test set and query budget as the main experiment, and were evaluated using TP-ASR@5, TP-ASR@10, AUC-TP and AvgQ.

\begin{table}[t]
  \centering
  \small
  \renewcommand{\arraystretch}{1.15}
  \setlength{\tabcolsep}{2mm}
  \begin{tabular}{lcccc}
    \hline
    \multirow{2}{*}{Variant} & \multicolumn{2}{c}{TP-ASR $\uparrow$} & \multirow{2}{*}{AUC-TP $\uparrow$} & \multirow{2}{*}{AvgQ $\downarrow$} \\
    \cline{2-3}
                             & $B{=}5$       & $B{=}10$              &                                    &                                    \\
    \hline
    w/o HCC                  & 45.2          & 61.1                  & 0.438                              & 6.6                                \\
    w/o EAVE                 & 42.0          & 57.8                  & 0.401                              & 7.0                                \\
    w/o BACR                 & 44.6          & 60.5                  & 0.423                              & 6.8                                \\
    \textbf{TempQ-Jail}      & \textbf{48.9} & \textbf{65.4}         & \textbf{0.469}                     & \textbf{6.3}                       \\
    \hline
  \end{tabular}
  \caption{Ablation study of the three core modules in TempQ-Jail.}
  \label{tab:ablation}
\end{table}

Two key observations can be drawn from Table~\ref{tab:ablation}: (1) All three core modules make a positive contribution to the final performance, with the end-to-end attack value estimation playing the most significant role. This indicates that, in query-constrained scenarios, accurately identifying candidates with full attack potential is key to improving the success rate of strict attacks. (2) Budget-aware ranking primarily enhances the forward accessibility of high-value candidates, whilst heterogeneous candidate construction improves the coverage of the candidate space. The two approaches enhance attack efficiency under limited budgets by addressing query priority and candidate supply, respectively.

\paragraph{Category-wise Performance.}
To analyse the performance of different attack methods across different security risk categories, we further compare the TP-ASR@5 scores across six risk categories. We report both the Micro Avg. and the Macro Avg.; the former calculates the overall TP-ASR@5 directly across all test intents and therefore takes into account the sample sizes of different categories; the latter is the equally weighted average of the TP-ASR@5 values across the six risk categories, used to measure overall performance across categories. The results are shown in Table~\ref{tab:category_performance}.

\begin{table*}[t]
  \centering
  \renewcommand{\arraystretch}{1.15}
  \setlength{\tabcolsep}{4mm}
  \begin{tabular}{lccccccc}
    \hline
    Category             & DACA          & T2V-OptJail   & SceneSplit    & SPARK         & TFM           & BSB           & TempQ-Jail    \\
    \hline
    Violence             & 49            & 47            & 21            & 19            & 13            & 24            & \textbf{57}   \\
    Gore                 & 7             & \textbf{41}   & 4             & 0             & 5             & 8             & 27            \\
    Self-harm            & 49            & 43            & 11            & 30            & 16            & 33            & \textbf{56}   \\
    Illegal Activity     & \textbf{66}   & 50            & 23            & 29            & 21            & 43            & 61            \\
    Disturbing Content   & \textbf{50}   & 40            & 6             & 6             & 14            & 28            & 38            \\
    Pornography          & 53            & 40            & 0             & 33            & 0             & 20            & \textbf{67}   \\
    \hline
    \textbf{Micro Avg.}  & 43.4          & 44.3          & 12.9          & 17.7          & 13.1          & 26.6          & \textbf{48.9} \\
    \textbf{Macro Avg.}  & 45.5          & 43.6          & 11.0          & 19.4          & 11.6          & 26.0          & \textbf{50.9} \\
    \hline
  \end{tabular}
  \caption{Category-wise TP-ASR@5 performance across six security risk categories.}
  \label{tab:category_performance}
\end{table*}

Two key observations can be drawn from Table~\ref{tab:category_performance}: (1) Different attack mechanisms exhibit distinct category preferences across different risk categories. TempQ-Jail achieves the highest TP-ASR@5 for ``Violence'', ``Self-harm'' and ``Pornography'', whilst T2V-OptJail performs best for ``Gore'', and DACA excels in ``Illegal Activity'' and ``Disturbing Content'', indicating that different candidate construction mechanisms possess varying degrees of adaptability to different hazardous semantics. (2) TempQ-Jail achieves the best results in terms of overall cross-category performance. Its micro average and macro average reach 48.9\% and 50.9\% respectively, both exceeding those of all comparison methods. This indicates that heterogeneous candidate construction combined with unified value ranking can comprehensively leverage the strengths of different attack mechanisms to achieve more stable overall attack performance across multiple risk categories.


\section{Conclusion}
\label{sec:conclusion}

This paper investigates the problem of query-constrained black-box jailbreak attacks in guarded T2V systems. Unlike existing work, which primarily focuses on generating more or more covert attack candidates, we further examine the practical accessibility of candidates under a limited target query budget and propose TempQ-Jail. This method expands the attack candidate space through heterogeneous candidate construction; employs end-to-end attack value estimation that comprehensively considers security gate passage, dangerous visual generation, preservation of original intent, and temporal validity; and utilises budget-aware candidate ranking to prioritise high-value candidates at the front of a finite query trajectory. In doing so, it further transforms candidate generation into a candidate allocation and ranking problem tailored to finite budgets.

Unified evaluations on CogVideoX-5B demonstrate that TempQ-Jail achieves the strongest overall attack performance across different query budgets, with TP-ASR@5 and TP-ASR@10 reaching 48.9\% and 65.4\%, representing improvements of 4.6 and 4.0 percentage points over the strongest baselines under the corresponding budgets, whilst achieving the highest AUC-TP (0.469) and the lowest AvgQ (6.3). Further analyses of query trajectories, candidate allocation, failure attribution and ablation studies demonstrate that TempQ-Jail is capable of more effectively identifying and prioritising candidates with full attack potential. Overall, this paper demonstrates that candidate generation capability and candidate accessibility under a limited budget are two distinct yet equally important issues in T2V red teaming, whilst TempQ-Jail further unifies candidate value assessment with query priority optimisation within a query-constrained attack framework.


\section*{Declarations}

\begin{small}

\noindent\textbf{Availability of data and material.}
The benchmark data used in this study are derived from the publicly available
T2VSafetyBench. Additional processed evaluation data supporting the findings of
this study are available from the corresponding author upon reasonable request.
Safety-sensitive attack materials may be subject to appropriate access
restrictions to mitigate potential misuse.

\vspace{.15in}
\noindent\textbf{Competing interests.}
The authors declare that they have no competing financial interests or personal
relationships that could have appeared to influence the work reported in this
paper.

\vspace{.15in}
\noindent\textbf{Funding.}
No funding was received for conducting this study.

\vspace{.15in}
\noindent\textbf{Authors' contributions.}
Conceptualization: Tianmeng Fang, Jiayang Liu, Xiaochun Cao;
Methodology: Tianmeng Fang, Jiancheng Wang, Chen Wang;
Software: Tianmeng Fang, Jiancheng Wang;
Formal analysis and investigation: Tianmeng Fang, Jiancheng Wang, Chen Wang;
Data curation: Jiancheng Wang, Liming Wang;
Visualization: Tianmeng Fang, Chen Wang;
Writing --- original draft preparation: Tianmeng Fang;
Writing --- review and editing: Chen Wang, Wei Wang, Jiayang Liu, Xiaochun Cao;
Resources: Liming Wang, Wei Wang, Xiaochun Cao;
Supervision: Jiayang Liu, Wei Wang, Xiaochun Cao.
All authors read and approved the final manuscript.

\vspace{.15in}
\noindent\textbf{Ethics approval.}
This study involves no human participants and no animals, so ethics approval was
not required.

\vspace{.15in}
\noindent\textbf{Ethics statement and responsible disclosure.}
This work is red-teaming research: its purpose is to expose a failure mode of
guarded text-to-video systems---that a limited target query budget, rather than
the existence of unsafe candidates, is what an attacker is actually constrained
by---so that defenders can account for it. We believe the defensive value of
reporting this behaviour outweighs the marginal uplift it offers an adversary,
since every attack mechanism we combine is already published.
All experiments were conducted offline on a locally hosted, open-source model
(CogVideoX-5B) behind a safety gate that we deployed ourselves; at no point did
we attack a third-party or commercial text-to-video service, and no live system
or its users were affected. The unsafe intents originate from T2VSafetyBench, an
existing public safety benchmark, and were not newly authored for this study.
Generated videos were viewed only by the authors, solely to score attack
outcomes, and were retained in restricted local storage; the paper reports
aggregate statistics only, and reproduces no unsafe prompt text or generated
frame. Release of the safety-sensitive materials is governed by the
availability statement above.
At the time of submission we had not yet contacted the developers of CogVideoX.
We intend to notify them of these findings before this work is published and
before any of the associated attack materials are released, so that the
weakness we report can be addressed ahead of wider disclosure.

\vspace{.15in}
\noindent\textbf{Acknowledgements.}
The authors thank the developers of CogVideoX and T2VSafetyBench for making
their models and benchmarks publicly available, which made this study possible.

\end{small}

\bibliographystyle{unsrt}
\bibliography{reference}

\begin{thebibliography}{10}

\bibitem{yang2025cogvideox}
Zhuoyi Yang, Jiayan Teng, Wendi Zheng, Ming Ding, Shiyu Huang, Jiazheng Xu, Yuanming Yang, Wenyi Hong, Xiaohan Zhang, Guanyu Feng, et~al.
\newblock Cogvideox: Text-to-video diffusion models with an expert transformer.
\newblock In {\em International Conference on Learning Representations}, volume 2025, pages 83048--83077, 2025.

\bibitem{chen2024videocrafter2}
Haoxin Chen, Yong Zhang, Xiaodong Cun, Menghan Xia, Xintao Wang, Chao Weng, and Ying Shan.
\newblock Videocrafter2: Overcoming data limitations for high-quality video diffusion models.
\newblock In {\em 2024 IEEE/CVF Conference on Computer Vision and Pattern Recognition (CVPR)}, pages 7310--7320. IEEE, 2024.

\bibitem{kong2024hunyuanvideo}
Weijie Kong, Qi~Tian, Zijian Zhang, Rox Min, Zuozhuo Dai, Jin Zhou, Jiangfeng Xiong, Xin Li, Bo~Wu, Jianwei Zhang, et~al.
\newblock Hunyuanvideo: A systematic framework for large video generative models.
\newblock {\em arXiv preprint arXiv:2412.03603}, 2024.

\bibitem{wan2025wan}
Team Wan, Ang Wang, Baole Ai, Bin Wen, Chaojie Mao, Chen-Wei Xie, Di~Chen, Feiwu Yu, Haiming Zhao, Jianxiao Yang, et~al.
\newblock Wan: Open and advanced large-scale video generative models.
\newblock {\em arXiv preprint arXiv:2503.20314}, 2025.

\bibitem{brooks2024video}
Tim Brooks, Bill Peebles, Connor Holmes, Will DePue, Yufei Guo, Leo Jing, David Schnurr, Joe Taylor, Troy Luhman, Eric Luhman, et~al.
\newblock Video generation models as world simulators.
\newblock {\em OpenAI Blog}, 1(8):1, 2024.

\bibitem{miao2024t2vsafetybench}
Yibo Miao, Yifan Zhu, Lijia Yu, Jun Zhu, Xiao-Shan Gao, and Yinpeng Dong.
\newblock T2vsafetybench: Evaluating the safety of text-to-video generative models.
\newblock {\em Advances in Neural Information Processing Systems}, 37:63858--63872, 2024.

\bibitem{lee2026jailbreaking}
Wonjun Lee, Haon Park, Doehyeon Lee, Bumsub Ham, and Suhyun Kim.
\newblock Jailbreaking on text-to-video models via scene splitting strategy.
\newblock In {\em International Conference on Learning Representations}, volume 2026, pages 58890--58920, 2026.

\bibitem{jing2026cogmorph}
Zonglei Jing, Zonghao Ying, Le~Wang, Siyuan Liang, Xiaoqian Li, Mingchuan Zhang, Aishan Liu, Xianglong Liu, and Dacheng Tao.
\newblock Cogmorph: Cognitive morphing attacks for text-to-image models.
\newblock {\em IEEE Transactions on Dependable and Secure Computing}, 2026.

\bibitem{ying2025pushing}
Zonghao Ying, Siyang Wu, Run Hao, Peng Ying, Shixuan Sun, Pengyu Chen, Junze Chen, Hao Du, Kaiwen Shen, Shangkun Wu, et~al.
\newblock Pushing the limits of safety: A technical report on the atlas challenge 2025.
\newblock {\em arXiv preprint arXiv:2506.12430}, 2025.

\bibitem{liang2026t2vshield}
Siyuan Liang, Jiayang Liu, Jiecheng Zhai, Tianmeng Fang, Rongcheng Tu, Aishan Liu, Xiaochun Cao, and Dacheng Tao.
\newblock T2vshield: Model-agnostic jailbreak defense for text-to-video models.
\newblock {\em International Journal of Computer Vision}, 134(4):144, 2026.

\bibitem{lu2025adversarial}
Liming Lu, Shuchao Pang, Siyuan Liang, Haotian Zhu, Xiyu Zeng, Aishan Liu, Yunhuai Liu, and Yongbin Zhou.
\newblock Adversarial training for multimodal large language models against jailbreak attacks.
\newblock {\em arXiv preprint arXiv:2503.04833}, 2025.

\bibitem{pang2026safesteer}
Shuchao Pang, Xiyu Zeng, Siyuan Liang, Liming Lu, Haotian Zhu, Chuanting Zhang, Enguang Liu, Liang Zhang, Basem Shihada, Yongbin Zhou, et~al.
\newblock Safesteer: Adaptive subspace steering for efficient jailbreak defense in vision language models.
\newblock {\em IEEE Transactions on Information Forensics and Security}, 2026.

\bibitem{jing2025promptsafe}
Zonglei Jing, Xiao Yang, Xiaoqian Li, Siyuan Liang, Aishan Liu, Mingchuan Zhang, and Xianglong Liu.
\newblock Promptsafe: Gated prompt tuning for safe text-to-image generation.
\newblock {\em arXiv preprint arXiv:2508.01272}, 2025.

\bibitem{xiao2025detoxifying}
Yisong Xiao, Aishan Liu, Siyuan Liang, Zonghao Ying, Xianglong Liu, and Dacheng Tao.
\newblock Detoxifying large language models via autoregressive reward guided representation editing.
\newblock {\em arXiv preprint arXiv:2510.01243}, 2025.

\bibitem{deng2023divide}
Yimo Deng and Huangxun Chen.
\newblock Divide-and-conquer attack: Harnessing the power of llm to bypass the censorship of text-to-image generation model.
\newblock {\em arXiv preprint arXiv:2312.07130}, 1(2):6, 2023.

\bibitem{liu2026t2v}
Jiayang Liu, Siyuan Liang, Shiqian Zhao, Rong-Cheng Tu, Wenbo Zhou, Aishan Liu, Dacheng Tao, and Siew~Kei Lam.
\newblock T2v-optjail: Discrete prompt optimization for text-to-video jailbreak attacks.
\newblock {\em Advances in Neural Information Processing Systems}, 38:73752--73770, 2026.

\bibitem{ying2025spark}
Zonghao Ying, Moyang Chen, Nizhang Li, Zhiqiang Wang, Wenxin Zhang, Quanchen Zou, Zonglei Jing, Aishan Liu, and Xianglong Liu.
\newblock Spark: Jailbreaking t2v models by synergistically prompting auditory and recontextualized knowledge.
\newblock {\em arXiv preprint arXiv:2511.13127}, 2025.

\bibitem{li2024semantic}
Xiaoxia Li, Siyuan Liang, Jiyi Zhang, Han Fang, Aishan Liu, and Ee-Chien Chang.
\newblock Semantic mirror jailbreak: Genetic algorithm based jailbreak prompts against open-source llms.
\newblock {\em arXiv preprint arXiv:2402.14872}, 2024.

\bibitem{ying2024jailbreak}
Zonghao Ying, Aishan Liu, Tianyuan Zhang, Zhengmin Yu, Siyuan Liang, Xianglong Liu, and Dacheng Tao.
\newblock Jailbreak vision language models via bi-modal adversarial prompt.
\newblock {\em arXiv preprint arXiv:2406.04031}, 2024.

\bibitem{ying2025reasoning}
Zonghao Ying, Deyue Zhang, Zonglei Jing, Yisong Xiao, Quanchen Zou, Aishan Liu, Siyuan Liang, Xiangzheng Zhang, Xianglong Liu, and Dacheng Tao.
\newblock Reasoning-augmented conversation for multi-turn jailbreak attacks on large language models.
\newblock {\em arXiv preprint arXiv:2502.11054}, 2025.

\bibitem{wang2025manipulating}
Le~Wang, Zonghao Ying, Tianyuan Zhang, Siyuan Liang, Shengshan Hu, Mingchuan Zhang, Aishan Liu, and Xianglong Liu.
\newblock Manipulating multimodal agents via cross-modal prompt injection.
\newblock In {\em Proceedings of the 33rd ACM International Conference on Multimedia}, pages 10955--10964, 2025.

\bibitem{chen2026two}
Moyang Chen, Zonghao Ying, Wenzhuo Xu, Quancheng Zou, Deyue Zhang, Dongdong Yang, and Xiangzheng Zhang.
\newblock Two frames matter: A temporal attack for text-to-video model jailbreaking.
\newblock {\em arXiv preprint arXiv:2603.07028}, 2026.

\bibitem{peng2026between}
Xingkai Peng, Jun Jiang, Jiayang Liu, Kejiang Chen, and Weiming Zhang.
\newblock Between safe boundaries: Exploiting temporal consistency for jailbreaking text-to-video generation models.
\newblock {\em arXiv preprint arXiv:2607.17279}, 2026.

\bibitem{chao2024jailbreakbench}
Patrick Chao, Edoardo Debenedetti, Alexander Robey, Maksym Andriushchenko, Francesco Croce, Vikash Sehwag, Edgar Dobriban, Nicolas Flammarion, George~J Pappas, Florian Tramer, et~al.
\newblock Jailbreakbench: An open robustness benchmark for jailbreaking large language models.
\newblock {\em Advances in Neural Information Processing Systems}, 37:55005--55029, 2024.

\bibitem{ho2022video}
Jonathan Ho, Tim Salimans, Alexey Gritsenko, William Chan, Mohammad Norouzi, and David~J Fleet.
\newblock Video diffusion models.
\newblock {\em Advances in neural information processing systems}, 35:8633--8646, 2022.

\bibitem{ho2022imagen}
Jonathan Ho, William Chan, Chitwan Saharia, Jay Whang, Ruiqi Gao, Alexey Gritsenko, Diederik~P Kingma, Ben Poole, Mohammad Norouzi, David~J Fleet, et~al.
\newblock Imagen video: High definition video generation with diffusion models.
\newblock {\em arXiv preprint arXiv:2210.02303}, 2022.

\bibitem{singer2022make}
Uriel Singer, Adam Polyak, Thomas Hayes, Xi~Yin, Jie An, Songyang Zhang, Qiyuan Hu, Harry Yang, Oron Ashual, Oran Gafni, et~al.
\newblock Make-a-video: Text-to-video generation without text-video data.
\newblock {\em arXiv preprint arXiv:2209.14792}, 2022.

\bibitem{blattmann2023align}
Andreas Blattmann, Robin Rombach, Huan Ling, Tim Dockhorn, Seung~Wook Kim, Sanja Fidler, and Karsten Kreis.
\newblock Align your latents: High-resolution video synthesis with latent diffusion models.
\newblock In {\em 2023 IEEE/CVF Conference on Computer Vision and Pattern Recognition (CVPR)}, pages 22563--22575. IEEE, 2023.

\bibitem{luo2023videofusion}
Zhengxiong Luo, Dayou Chen, Yingya Zhang, Yan Huang, Liang Wang, Yujun Shen, Deli Zhao, Jingren Zhou, and Tieniu Tan.
\newblock Videofusion: Decomposed diffusion models for high-quality video generation.
\newblock {\em arXiv preprint arXiv:2303.08320}, 2023.

\bibitem{khachatryan2023text2video}
Levon Khachatryan, Andranik Movsisyan, Vahram Tadevosyan, Roberto Henschel, Zhangyang Wang, Shant Navasardyan, and Humphrey Shi.
\newblock Text2video-zero: Text-to-image diffusion models are zero-shot video generators.
\newblock In {\em 2023 IEEE/CVF International Conference on Computer Vision (ICCV)}, pages 15908--15918. IEEE, 2023.

\bibitem{guo2023animatediff}
Yuwei Guo, Ceyuan Yang, Anyi Rao, Zhengyang Liang, Yaohui Wang, Yu~Qiao, Maneesh Agrawala, Dahua Lin, and Bo~Dai.
\newblock Animatediff: Animate your personalized text-to-image diffusion models without specific tuning.
\newblock {\em arXiv preprint arXiv:2307.04725}, 2023.

\bibitem{ma2024latte}
Xin Ma, Yaohui Wang, Xinyuan Chen, Gengyun Jia, Ziwei Liu, Yuan-Fang Li, Cunjian Chen, and Yu~Qiao.
\newblock Latte: Latent diffusion transformer for video generation.
\newblock {\em arXiv preprint arXiv:2401.03048}, 2024.

\bibitem{bar2024lumiere}
Omer Bar-Tal, Hila Chefer, Omer Tov, Charles Herrmann, Roni Paiss, Shiran Zada, Ariel Ephrat, Junhwa Hur, Guanghui Liu, Amit Raj, et~al.
\newblock Lumiere: A space-time diffusion model for video generation.
\newblock In {\em SIGGRAPH Asia 2024 conference papers}, pages 1--11, 2024.

\bibitem{wang2023modelscope}
Jiuniu Wang, Hangjie Yuan, Dayou Chen, Yingya Zhang, Xiang Wang, and Shiwei Zhang.
\newblock Modelscope text-to-video technical report.
\newblock {\em arXiv preprint arXiv:2308.06571}, 2023.

\bibitem{zheng2024open}
Zangwei Zheng, Xiangyu Peng, Tianji Yang, Chenhui Shen, Shenggui Li, Hongxin Liu, Yukun Zhou, Tianyi Li, and Yang You.
\newblock Open-sora: Democratizing efficient video production for all.
\newblock {\em arXiv preprint arXiv:2412.20404}, 2024.

\bibitem{li2026sok}
Siyuan Li, Aodu Wulianghai, Zehao Liu, Xi~Lin, Qinghua Mao, Haoyu Li, Xiang Chen, Siyuan Liang, Jun Wu, Jianhua Li, et~al.
\newblock Sok: Intent-oriented systematization of multi-turn llm jailbreaks.
\newblock {\em arXiv preprint arXiv:2608.01117}, 2026.

\bibitem{yang2024sneakyprompt}
Yuchen Yang, Bo~Hui, Haolin Yuan, Neil Gong, and Yinzhi Cao.
\newblock Sneakyprompt: Jailbreaking text-to-image generative models.
\newblock In {\em 2024 IEEE symposium on security and privacy (SP)}, pages 897--912. IEEE, 2024.

\bibitem{liang2022parallel}
Siyuan Liang, Baoyuan Wu, Yanbo Fan, Xingxing Wei, and Xiaochun Cao.
\newblock Parallel rectangle flip attack: A query-based black-box attack against object detection.
\newblock {\em arXiv preprint arXiv:2201.08970}, 2022.

\bibitem{liang2022large}
Siyuan Liang, Longkang Li, Yanbo Fan, Xiaojun Jia, Jingzhi Li, Baoyuan Wu, and Xiaochun Cao.
\newblock A large-scale multiple-objective method for black-box attack against object detection.
\newblock In {\em European Conference on Computer Vision}, 2022.

\bibitem{liang2025hard}
Muxue Liang, Chuan Wang, Siyuan Liang, Aishan Liu, Yanan Cao, Qingyong Li, Zeming Liu, Liang Yang, and Xiaochun Cao.
\newblock Hard-label black-box adversarial attacks for implicit scene interactions.
\newblock {\em IEEE Transactions on Information Forensics and Security}, 20:10346--10360, 2025.

\bibitem{wang2025black}
Lu~Wang, Tianyuan Zhang, Yang Qu, Siyuan Liang, Yuwei Chen, Aishan Liu, Xianglong Liu, and Dacheng Tao.
\newblock Black-box adversarial attack on vision language models for autonomous driving.
\newblock {\em arXiv preprint arXiv:2501.13563}, 2025.

\bibitem{wang2025no}
Wenqiang Wang, Siyuan Liang, Yangshijie Zhang, Xiaojun Jia, Hao Lin, and Xiaochun Cao.
\newblock No query, no access.
\newblock {\em arXiv preprint arXiv:2505.07258}, 2025.

\bibitem{tsai2024ring}
Yu-Lin Tsai, Chia-Yi Hsu, Chulin Xie, Chih-Hsun Lin, Jia~You Chen, Bo~Li, Pin-Yu Chen, Chia-Mu Yu, and Chun-Ying Huang.
\newblock Ring-a-bell! how reliable are concept removal methods for diffusion models?
\newblock In {\em International Conference on Learning Representations}, volume 2024, pages 41543--41554, 2024.

\bibitem{chin2023prompting4debugging}
Zhi-Yi Chin, Chieh-Ming Jiang, Ching-Chun Huang, Pin-Yu Chen, and Wei-Chen Chiu.
\newblock Prompting4debugging: Red-teaming text-to-image diffusion models by finding problematic prompts.
\newblock {\em arXiv preprint arXiv:2309.06135}, 2023.

\bibitem{guo2026wmattack}
Zhixiang Guo, Siyuan Liang, Shi Fu, Cheng Guo, Andras Balogh, Mark Jelasity, and Dacheng Tao.
\newblock Wmattack: Automated attack search for adversarial evaluation of world-model agents.
\newblock {\em arXiv preprint arXiv:2605.23220}, 2026.

\bibitem{xu2026ctrlattack}
Shuhan Xu, Siyuan Liang, Hongling Zheng, Yong Luo, Han Hu, Lefei Zhang, and Dacheng Tao.
\newblock Ctrlattack: A unified attack on world-model control in diffusion models.
\newblock {\em arXiv preprint arXiv:2603.13435}, 2026.

\bibitem{wang2026text}
Wenqiang Wang, Siyuan Liang, Yangshijie Zhang, Zhifeng Chen, Yan Xiao, and Xiaochun Cao.
\newblock Text adversarial attacks with dynamic outputs.
\newblock {\em IEEE Transactions on Information Forensics and Security}, 2026.

\bibitem{yang2024mma}
Yijun Yang, Ruiyuan Gao, Xiaosen Wang, Tsung-Yi Ho, Nan Xu, and Qiang Xu.
\newblock Mma-diffusion: Multimodal attack on diffusion models.
\newblock In {\em Proceedings of the IEEE/CVF conference on computer vision and pattern recognition}, pages 7737--7746, 2024.

\bibitem{he2023sa}
Bangyan He, Xiaojun Jia, Siyuan Liang, Tianrui Lou, Yang Liu, and Xiaochun Cao.
\newblock Sa-attack: Improving adversarial transferability of vision-language pre-training models via self-augmentation.
\newblock {\em arXiv preprint arXiv:2312.04913}, 2023.

\bibitem{liu2025bridging}
Kuanrong Liu, Siyuan Liang, Cheng Qian, Ming Zhang, and Xiaochun Cao.
\newblock Bridging the task gap: Multi-task adversarial transferability in clip and its derivatives.
\newblock In {\em Chinese Conference on Pattern Recognition and Computer Vision (PRCV)}, pages 152--166. Springer, 2025.

\bibitem{zhang2026visual}
Tianyuan Zhang, Lu~Wang, Xinwei Zhang, Yitong Zhang, Boyi Jia, Siyuan Liang, Shengshan Hu, Qiang Fu, Aishan Liu, and Xianglong Liu.
\newblock Visual adversarial attack on vision-language models for autonomous driving.
\newblock {\em Machine Intelligence Research}, pages 1--18, 2026.

\bibitem{kong2025universal}
Dehong Kong, Sifan Yu, Siyuan Liang, Jiawei Liang, Jianhou Gan, Aishan Liu, and Wenqi Ren.
\newblock Universal camouflage attack on vision-language models for autonomous driving.
\newblock {\em arXiv preprint arXiv:2509.20196}, 2025.

\bibitem{kong2024patch}
Dehong Kong, Siyuan Liang, Xiaopeng Zhu, Yuansheng Zhong, and Wenqi Ren.
\newblock Patch is enough: naturalistic adversarial patch against vision-language pre-training models.
\newblock {\em Visual Intelligence}, 2(1):1--10, 2024.

\bibitem{liu2023improving}
Jiayang Liu, Siyu Zhu, Siyuan Liang, Jie Zhang, Han Fang, Weiming Zhang, and Ee-Chien Chang.
\newblock Improving adversarial transferability by stable diffusion.
\newblock {\em arXiv preprint arXiv:2311.11017}, 2023.

\bibitem{ma2025jailbreaking}
Jiachen Ma, Yijiang Li, Zhiqing Xiao, Anda Cao, Jie Zhang, Chao Ye, and Junbo Zhao.
\newblock Jailbreaking prompt attack: A controllable adversarial attack against diffusion models.
\newblock In {\em Findings of the Association for Computational Linguistics: NAACL 2025}, pages 3141--3157, 2025.

\bibitem{liu2025multimodal}
Tong Liu, Zhixin Lai, Jiawen Wang, Gengyuan Zhang, Shuo Chen, Philip Torr, Vera Demberg, Volker Tresp, and Jindong Gu.
\newblock Multimodal pragmatic jailbreak on text-to-image models.
\newblock In {\em Proceedings of the 63rd Annual Meeting of the Association for Computational Linguistics (Volume 1: Long Papers)}, pages 4681--4720, 2025.

\bibitem{jin2025jailbreakdiffbench}
Xiaolong Jin, Zixuan Weng, Hanxi Guo, Chenlong Yin, Siyuan Cheng, Guangyu Shen, and Xiangyu Zhang.
\newblock Jailbreakdiffbench: A comprehensive benchmark for jailbreaking diffusion models.
\newblock In {\em 2025 IEEE/CVF International Conference on Computer Vision (ICCV)}, pages 16461--16471. IEEE, 2025.

\bibitem{ying2026safebench}
Zonghao Ying, Aishan Liu, Siyuan Liang, Lei Huang, Jinyang Guo, Wenbo Zhou, Xianglong Liu, and Dacheng Tao.
\newblock Safebench: A safety evaluation framework for multimodal large language models.
\newblock {\em International Journal of Computer Vision}, 134(1):18, 2026.

\bibitem{liu2025agentsafe}
Aishan Liu, Zonghao Ying, Le~Wang, Junjie Mu, Jinyang Guo, Jiakai Wang, Yuqing Ma, Siyuan Liang, Mingchuan Zhang, Xianglong Liu, et~al.
\newblock Agentsafe: Benchmarking the safety of embodied agents on hazardous instructions.
\newblock {\em arXiv preprint arXiv:2506.14697}, 2025.

\bibitem{zhang2025bench2advlm}
Tianyuan Zhang, Ting Jin, Lu~Wang, Jiangfan Liu, Siyuan Liang, Mingchuan Zhang, Aishan Liu, and Xianglong Liu.
\newblock Bench2advlm: a closed-loop benchmark for vision-language models in autonomous driving.
\newblock {\em arXiv preprint arXiv:2508.02028}, 2025.

\bibitem{liu2025metadv}
Aishan Liu, Jiakai Wang, Tianyuan Zhang, Hainan Li, Jiangfan Liu, Siyuan Liang, Yilong Ren, Xianglong Liu, and Dacheng Tao.
\newblock Metadv: A unified and interactive adversarial testing platform for autonomous driving.
\newblock In {\em Proceedings of the 33rd ACM International Conference on Multimedia}, pages 13474--13476, 2025.

\bibitem{dai2024safesora}
Juntao Dai, Tianle Chen, Xuyao Wang, Ziran Yang, Taiye Chen, Jiaming Ji, and Yaodong Yang.
\newblock Safesora: Towards safety alignment of text2video generation via a human preference dataset.
\newblock {\em Advances in neural information processing systems}, 37:17161--17214, 2024.

\bibitem{huang2024vbench}
Ziqi Huang, Yinan He, Jiashuo Yu, Fan Zhang, Chenyang Si, Yuming Jiang, Yuanhan Zhang, Tianxing Wu, Qingyang Jin, Nattapol Chanpaisit, et~al.
\newblock Vbench: Comprehensive benchmark suite for video generative models.
\newblock In {\em 2024 IEEE/CVF Conference on Computer Vision and Pattern Recognition (CVPR)}, pages 21807--21818. IEEE, 2024.

\bibitem{liu2024evalcrafter}
Yaofang Liu, Xiaodong Cun, Xuebo Liu, Xintao Wang, Yong Zhang, Haoxin Chen, Yang Liu, Tieyong Zeng, Raymond Chan, and Ying Shan.
\newblock Evalcrafter: Benchmarking and evaluating large video generation models.
\newblock In {\em 2024 IEEE/CVF Conference on Computer Vision and Pattern Recognition (CVPR)}, pages 22139--22149. IEEE, 2024.

\bibitem{ma2026safegen}
Yingzi Ma, Xiaogeng Liu, Yawen Zheng, and Chaowei Xiao.
\newblock Safegen-bench: Benchmarking safety in image-conditioned text-to-video generation.
\newblock {\em arXiv preprint arXiv:2606.01481}, 2026.

\bibitem{yang2024guardt2i}
Yijun Yang, Ruiyuan Gao, Xiao Yang, Jianyuan Zhong, and Qiang Xu.
\newblock Guardt2i: Defending text-to-image models from adversarial prompts.
\newblock {\em Advances in neural information processing systems}, 37:76380--76403, 2024.

\bibitem{schramowski2023safe}
Patrick Schramowski, Manuel Brack, Bj{\"o}rn Deiseroth, and Kristian Kersting.
\newblock Safe latent diffusion: Mitigating inappropriate degeneration in diffusion models.
\newblock In {\em 2023 IEEE/CVF Conference on Computer Vision and Pattern Recognition (CVPR)}, pages 22522--22531. IEEE, 2023.

\bibitem{gandikota2023erasing}
Rohit Gandikota, Joanna Materzynska, Jaden Fiotto-Kaufman, and David Bau.
\newblock Erasing concepts from diffusion models.
\newblock In {\em Proceedings of the IEEE/CVF international conference on computer vision}, pages 2426--2436, 2023.

\bibitem{wang2026runawayevil}
Songping Wang, Rufan Qian, Yueming Lyu, Qinglong Liu, Linzhuang Zou, Jie Qin, Songhua Liu, and Caifeng Shan.
\newblock Runawayevil: Jailbreaking the image-to-video generative models.
\newblock In {\em Proceedings of the IEEE/CVF Conference on Computer Vision and Pattern Recognition}, pages 9296--9305, 2026.

\bibitem{liang2025safemobile}
Siyuan Liang, Tianmeng Fang, Zhe Liu, Aishan Liu, Yan Xiao, Jinyuan He, Ee-Chien Chang, and Xiaochun Cao.
\newblock Safemobile: Chain-level jailbreak detection and automated evaluation for multimodal mobile agents.
\newblock {\em arXiv preprint arXiv:2507.00841}, 2025.

\end{thebibliography}

\end{document}